\documentclass[%
 aip,
 amsmath,amssymb,
 reprint,%
]{revtex4-1}

\usepackage[english]{babel}

\makeatletter
\@namedef{l@en}{\l@english}
\@namedef{captionsen}{\captionsenglish}
\@namedef{dateen}{\dateenglish}
\@namedef{extrasen}{\extrasenglish}
\@namedef{noextrasen}{\noextrasenglish}
\makeatother

\usepackage{graphicx}
\usepackage{dcolumn}
\usepackage{bm}

\usepackage[utf8]{inputenc}
\usepackage[T1]{fontenc}
\usepackage{mathptmx}
\usepackage{etoolbox}
\usepackage{hyperref}

\hypersetup{
    colorlinks=true,
    linkcolor=black,
    urlcolor=black,
    citecolor=black
}

\usepackage{booktabs}
\usepackage{dsfont}

\begin{document}


\title{Correlation-Free Transition Path Sampling through Shooting Point Generation Guided by Committor Learning} 



\author{Maximilian Negedly}
\affiliation{Faculty of Physics, University of Vienna, 1090 Vienna, Austria}
\affiliation{Vienna Doctoral School in Physics, University of Vienna, 1090 Vienna, Austria}

\author{Sebastian Falkner}
\affiliation{Faculty of Physics, University of Vienna, 1090 Vienna, Austria}
\affiliation{Institute of Physics, University of Augsburg, 86159 Augsburg, Germany}

\author{Alessandro Coretti}
\affiliation{Faculty of Physics, University of Vienna, 1090 Vienna, Austria}

\author{Christoph Dellago} %
 \email{christoph.dellago@univie.ac.at}
\affiliation{Faculty of Physics, University of Vienna, 1090 Vienna, Austria}


\date{\today}

\begin{abstract}
    Studying the dynamical behavior of a system often depends on characterizing how it transitions between long-lived states. Because such transitions are rare, observing them usually requires specialized enhanced sampling techniques. Transition Path Sampling (TPS) is a well-established method for generating reactive trajectories, which is simple to implement and does not require the definition of a preconceived reaction coordinate. However, its efficiency is limited by its sequential nature and the resulting correlations between sampled paths. Previous work addressed this limitation by combining TPS with a sampling scheme based on conditioned Boltzmann Generators, a generative machine learning model capable of sampling a given target probability distribution. This approach produces uncorrelated transition paths but relies on an accurate reaction coordinate, which is rarely known in advance. Building on recent advances in committor learning, specifically on the Artificial Intelligence for Molecular Mechanism Discovery (AIMMD) method, in this work we introduce GenAIMMD, an iterative algorithm that actively and self-consistently learns the ideal reaction coordinate (the committor) and trains a conditioned Boltzmann Generator to sample from arbitrary bias windows along it. GenAIMMD thereby provides a correlation-free and fully parallelizable path sampling scheme that does not require prior knowledge of the system's transition mechanism. We apply GenAIMMD to a two-dimensional toy model and a higher-dimensional polymer system. In both cases, GenAIMMD succeeds in training the Boltzmann Generator and learning the committor. Benchmark results show a substantial increase in performance compared to standard TPS.
\end{abstract}


\maketitle

\section{Introduction}

Rare events, which occur in many processes ranging from chemical reactions to phase transitions, are difficult to study with conventional simulation methods, because their characteristic waiting times often exceed accessible simulation time scales.\cite{pan_dynamics_2004,juraszek_sampling_2006,okazaki_mechanism_2019,angiolari_electrically_2025,falkner_kinetic_2021} Established enhanced sampling approaches address this rare event problem either in configuration space through biasing techniques\cite{henin_enhanced_2022} such as Umbrella Sampling~\cite{torrie_nonphysical_1977, kastner_umbrella_2011} and Metadynamics,\cite{laio_escaping_2002} or in trajectory space through path-based methods~\cite{dellago_transition_2009} such as Transition Path Sampling (TPS). Despite their success, these approaches have important limitations. Their sequential nature produces correlated samples and restricts parallelization, while many methods additionally require the definition of suitable reaction coordinates (RCs). These limitations can lead to substantial computational costs, particularly for complex systems with poorly understood transition mechanisms.

Recent advances in machine learning (ML) for rare-event sampling, particularly for collective variables, committor learning~\cite{ma_automatic_2005, peters_obtaining_2006, jung_machine-guided_2023,kang_computing_2024,megias_iterative_2025} and generative models,\cite{falkner_conditioning_2023,asghar_efficient_2024,li_differentiable_2026,tang_breaking_2026} promise to help overcome some of these limitations. However, they remain fragmented and lack integration into a unified framework. Moreover, important challenges persist. Committor-learning methods typically rely on reactive trajectories, whose generation is computationally expensive and inherits some of the limitations of TPS, including strong correlations between successively sampled pathways that hinder the exploration of multiple reaction mechanisms. Conversely, the use of generative models in the context of rare events requires conditioning to target the transition-state region, potentially reintroducing the need for predefined reaction coordinates.

In this work, we develop an active learning framework that integrates Artificial Intelligence for Molecular Mechanism Discovery (AIMMD)~\cite{jung_machine-guided_2023} with conditioned Boltzmann Generators (cBGs) for rare-event sampling~\cite{falkner_conditioning_2023} within a unified simulation scheme. Reflecting this combination of generative sampling and AIMMD, we call the framework GenAIMMD. Starting from an initial set of training data generated using a combination of equilibrium and path sampling techniques, GenAIMMD learns a committor model that identifies the transition region and conditions the generator to produce configurations likely to initiate reactive trajectories. The active learning loop then iteratively refines both models, without further Markov-chain-based trajectory generation. By eliminating the need for predefined RCs and generating independent shooting points, this integrated approach removes correlations and enables the efficient parallel generation of independent trajectories while providing access to both thermodynamic and dynamical information. 

The remainder of the paper is organized as follows. In Sec.~\ref{sec:methods}, we review the three building blocks underlying our approach, namely TPS, AIMMD, and cBGs, and describe their integration into the GenAIMMD iterative algorithm. In Sec.~\ref{sec:results}, we present numerical results for a two-dimensional model and a polymer model, and benchmark our method against standard TPS. Finally, in Sec.~\ref{sec:discussion}, we discuss limitations of the presented algorithm as well as possible future directions.
\section{Methods}
\label{sec:methods}

\subsection{Transition Path Sampling}
\label{sec:tps}

Like many other enhanced sampling schemes, TPS~\cite{dellago_transition_1998, bolhuis_transition_2002} addresses the time-scale separation problem that prevents standard methods from efficiently crossing large energetic or entropic barriers in configuration space. Specifically, TPS samples the equilibrium ensemble of transition paths, that is, the ensemble of trajectories that connect two disconnected regions A and B in configuration space, typically corresponding to stable or metastable states of the system. By sampling directly in the space of reactive trajectories, TPS avoids the long waiting periods between rare transitions encountered in unbiased simulations.

In essence, TPS is a Markov Chain Monte Carlo method that operates in the space of equilibrium transition paths, proposing new path candidates and accepting or rejecting them according to a criterion derived from detailed balance. The most widely used algorithm implementing this idea is the \textit{shooting algorithm}.\cite{dellago_efficient_1998} Starting from a current path $X$, a configuration (the shooting point) along that path is selected with a certain probability and, if necessary, perturbed. From this configuration, two new trajectories are generated, one forward and one backward in time. The latter is obtained by propagating the system with inverted momenta. The two trajectories are then joined together preserving continuity of positions and velocities, forming a new candidate path $X'$. If accepted, $X'$ forms the next state in the Markov chain.

Employing an appropriate acceptance criterion ensures that TPS samples the same ensemble of transition paths that would be obtained from a long equilibrium simulation. The acceptance criterion follows from the detailed balance condition in path space and has been shown for different algorithms~\cite{jung_transition_2017} to depend on the probability of selecting a shooting point $x_s$ on the path $X$. For the two-way shooting algorithm considered here, it can be written as
\begin{equation}
\label{eq:acc_general}
P_{\text{acc}}[X\to X'] = H_{\mathrm{AB}}[X']\min\left\{1, \frac{p_{\text{sel}}[x'_{s'}|X']}{p_{\text{sel}}[x_s|X]}\right\},
\end{equation}
where $p_{\text{sel}}[x_s|X]$ is the shooting point selection probability, and $H_{\mathrm{AB}}(X)$ is equal to 1 if $X$ is a valid reactive path and 0 otherwise. This acceptance criterion assumes an unperturbed shooting point, which can sensibly only be used in combination with a stochastic kernel for the generation of trajectories. For uniform shooting point selection, i.e., $p_{\text{sel}}[x_s|X(L)] = L^{-1}$, the acceptance criterion for paths of variable length reduces to~\cite{bolhuis_transition_2003,bolhuis_transition-path_2003}
\begin{equation}
\label{eq:acc_uniform}
P_{\text{acc}}[X(L)\to X'(L')] = H_{\mathrm{AB}}[X'(L')]\min\left\{1,\frac{L}{L'}\right\}.
\end{equation}

The main advantage of uniform shooting point selection is that it requires no prior information about the system under investigation. Its drawback, however, is that it can result in very low acceptance probabilities. This is because many proposed paths will result in excursions that begin and end in the same state rather than genuine transition pathways. One way to increase the acceptance rate in TPS is therefore to bias the shooting point selection probability toward the barrier region,\cite{jung_transition_2017, menzl_s-shooting_2016, bolhuis_transition_2021} for example by restricting shooting points to lie only within a prescribed range or by using different shooting point distributions centered near the barrier region. 

Although such biasing can increase the acceptance rate, it reintroduces a key limitation that is absent when an unbiased (e.g. uniform) shooting point distribution is used: the need for prior knowledge of the system, typically in the form of a suitable reaction coordinate. While this information can be inferred for a specific system by combining physical intuition with computational methods, a suitable reaction coordinate is, in general, not easy to identify. Moreover, it can depend on several variables specific to the simulation, such as the potential energy surface, the definition of the stable states, and the particular model used to describe the dynamics. Ideally, the optimal reaction coordinate would be given in form of the \emph{committor}~\cite{onsager_initial_1938, du_transition_1998, hummer_transition_2004, e_transition-path_2010} $p_{\mathrm{B}}(x)$, defined as the probability that a trajectory initiated from a configuration $x$ reaches the product state (i.e., $\mathrm{B}$) before reaching the reactant state (i.e., $\mathrm{A}$). Unfortunately, closed-form expressions for the committor are available only for a limited number of simple systems. Machine learning techniques therefore provide a promising approach to estimating the committor for more complex systems, as discussed in the next section.

\subsection{AIMMD}

The committor is a function defined on the full phase space and is analytically intractable, while its numerical estimation is generally computationally expensive. Nevertheless, knowledge of the committor is highly desirable, as it provides detailed insight into reaction mechanisms and represents the ideal reaction coordinate. In the context of TPS, the committor can be used to increase the probability of generating reactive path candidates in two-way shooting moves by biasing the selection of shooting points towards configurations with $p_{\mathrm{B}} \approx 0.5$. This substantially reduces the number of non-reactive paths.

This idea forms the basis of the Artificial Intelligence for Molecular Mechanism Discovery (AIMMD) method.\cite{jung_machine-guided_2023} AIMMD introduces a machine learning framework to reconstruct the committor from information obtained during a standard TPS simulation. The resulting committor estimate is then used to improve shooting point selection, leading to a progressively more efficient sampling procedure.

In the AIMMD framework, the committor is represented as
\begin{equation}
    p_{\mathrm{B}}(x|\theta) = \frac{1}{1 + e^{-q(x | \theta)}},
    \label{eq:p_of_q}
\end{equation}
where $q(x | \theta)$ is a multilayer perceptron with parameters $\theta$. To train this model, one initiates $N$ trajectories from each of $k$ shooting points $\{x_i\}_{i = 1 \dots k}$. For every shooting point $x_i$, the numbers of trajectories reaching states A and B are recorded as $n_{\mathrm{A}}(i)$ and $n_{\mathrm{B}}(i)$, respectively. Since each trajectory constitutes a Bernoulli trial, these outcomes follow the binomial probability~\cite{jung_machine-guided_2023}
\begin{equation}
    p(n_{\mathrm{A}}, n_{\mathrm{B}} | x_i) = {n_{\mathrm{A}} + n_{\mathrm{B}} \choose n_\mathrm{A}} [1 - p_\mathrm{B}(x_i|\theta)]^{n_\mathrm{A}} [p_\mathrm{B}(x_i|\theta)]^{n_\mathrm{B}}
    \label{eq:binomial_probability}
\end{equation}
and combining the probabilities for the $k$ shooting points yields the likelihood function
\begin{equation}
    \mathcal{L} = \prod_{i=1}^k p(n_{\mathrm{A}}(i), n_{\mathrm{B}}(i) | x_i).
    \label{eq:likelihood}
\end{equation}
The negative log-likelihood is then used as a loss function, which, using Eqs.~\eqref{eq:p_of_q} and~\eqref{eq:binomial_probability}, can be expressed as
\begin{equation}
    \begin{split}
        L_{\mathrm{AIMMD}} &= - \log \mathcal{L} \\
        &= \sum_{i = 1}^k \sum_{j = 1}^N \log \left( 1 + e^{s_{ij} q(x_i | \theta)} \right)\ +\ \mathrm{const.}
    \end{split}
\end{equation}
Here, the matrix $s \in M_{k \times N}(\{-1, 1\})$ encodes the outcome of the $j$-th shooting move from shooting point $x_i$, setting $s_{ij} = -1$ if state B was reached before state A and $s_{ij} = 1$ otherwise.

\subsection{Boltzmann Generators}
\label{sec:boltzmann_generators}

Boltzmann Generators~\cite{noe_boltzmann_2019} are generative machine learning models belonging to the broader class of normalizing flows.\cite{tabak_density_2010, tabak_family_2013, dinh_density_2017} Given a target distribution of the form $p_X(x) \propto \exp[- \beta U(x)]$ with $\beta = (k_\mathrm{B}T)^{-1}$ and $x \in \Omega_X$ and, optionally, a small set of independent samples from that distribution, normalizing flows parametrize the diffeomorphism $F_{zx} : \Omega_Z \to \Omega_X$ between the latent space $\Omega_Z$ of a simple, readily sampled prior distribution $p_Z(z)$ (e.g. uniform or Gaussian) and the configuration space $\Omega_X$ of the target distribution. Once trained, independent latent space samples $z \sim p_Z(z)$ can be transformed using $F_{zx}$ to generate uncorrelated samples from the target distribution at the computational cost of a forward pass through the network. For this reason, Boltzmann Generators have found broad applicability across different areas of statistical physics,\cite{coretti_boltzmann_2024} with notable applications for atomistic simulations in free-energy calculations,\cite{wirnsberger_targeted_2020,ahmad_free_2022,schebek_efficient_2024,li_differentiable_2026} equilibrium simulations of solid\cite{wirnsberger_normalizing_2021} and liquid systems\cite{jung_normalizing_2024,coretti_learning_2025} and rare event sampling.\cite{falkner_conditioning_2023,asghar_efficient_2024}

The form of $F_{zx}$ and its inverse $F_{zx}^{-1} := F_{xz}$ is chosen such that their Jacobian determinants --- $\det J_{zx}(z)$ and $\det J_{xz}(x)$, respectively --- are tractable and efficient to compute. Following the original design of Boltzmann Generators,\cite{noe_boltzmann_2019} in this work the Real NVP architecture~\cite{dinh_density_2017} is used, which splits the input into two channels and passes information between them through a series of affine transformations with learnable parameters $\varphi$ modeled as multilayer perceptrons, placing the model into the category of split coupling flows. A multivariate normal distribution is chosen as the prior distribution $p_Z(z)$.

Training a Boltzmann Generator combines two complementary strategies: \textit{training by energy} and \textit{training by example}. In both cases, samples are passed through the network and used to estimate the Kullback-Leibler (KL) divergence between the target distribution $p_\alpha(a)$ and the distribution $q_\alpha(a)$ produced by the Boltzmann Generator, where $\alpha$ and $a$ are placeholders for the space in which the KL divergence is calculated --- latent space ($Z$ and $z$) when training by energy and configuration space ($X$ and $x$) when training by example. The objective of training is then to minimize this quantity, which, combining the two paradigms by summing their respective KL divergence estimators, can be written as a loss function~\cite{noe_boltzmann_2019}
\begin{equation}
\begin{split}
    L_{\mathrm{BG}} = &\\
                    = \, \lambda_{\mathrm{KL}} \, &\mathbb{E}_{z \sim p_Z(z)} \left[ \beta U(F_{zx}(z|\varphi)) - \log (|\det J_{zx}(z|\varphi)|) \right] \, + \\
                    + \, \lambda_{\mathrm{ML}} \, &\mathbb{E}_{x \sim p_X(x)} \left[ \frac{1}{2} ||F_{xz}(x|\varphi)||^2 - \log(|\det J_{xz}(x|\varphi)|) \right],
\end{split}
\end{equation}
where $\lambda_{\mathrm{KL}}$, $\lambda_{\mathrm{ML}}$ control the strength of the contribution of the individual estimators.

The difference between the two paradigms lies in the direction in which the data flows through the generator --- $Z \to X$ for training by energy and $X \to Z$ for training by example. In the latter case, samples from the target distribution are required, which help converge the Boltzmann Generator especially during the early stages of training and avoiding mode collapse.\cite{nicoli_detecting_2023} Both strategies complement each other, balancing exploration with numerical stability and training efficiency.

Since normalizing flows are exact-likelihood models, the probability density of generated samples $q_X(x)$ is known analytically. This allows the exact computation of the importance weights of samples $x \sim q_X\bigl(F_{zx}(z)\bigr)$ with respect to the true target distribution~\cite{noe_boltzmann_2019} $p_X(x)$ given the original sample $z\sim p_Z(z)$ using
\begin{equation}
\begin{split}
    \omega(F_{zx}(z)) &= \frac{p_X(F_{zx}(z))}{q_X(F_{zx}(z))} \\
    &\propto e^{-\beta U(F_{zx}(z)) - \log p_Z(z) + \log(|\det J_{zx}(z)|)}.
    \label{eq:boltzmann_weights}
\end{split}
\end{equation}
Provided the learned distribution $q_X\bigl(F_{zx}(z)\bigr)$ has sufficient overlap with the target distribution $p_X(x)$, the latter can be recovered by re-weighting the generated samples.

The importance weights also provide information about the quality of the sampling. A quantitative measure of this quality is given by the Effective Sample Size (ESS),\cite{kish_survey_1965} which becomes the Relative Effective Sample Size (RESS)
\begin{equation}
    \mathrm{RESS} = \frac{1}{N} \frac{\left(\sum_{i=1}^{N} \omega_i \right)^2}{\sum_{i=1}^{N} \omega_i^2}
    \label{eq:ress}
\end{equation}
after dividing by $N$. The RESS is a number between $N^{-1}$ and one, equaling one if the sampling is perfect (i.e., if $\omega(x) = 1/N$ for all $N$ samples) and approaching zero in the opposite case.

\subsection{Conditioning Boltzmann Generators for TPS}

In the context of rare event sampling, it is desirable to bias the sample generation towards regions in configuration space that are visited during transitions but rarely in unbiased simulations. Such regions are typically characterized using a collective variable $\xi(x)$. Falkner et al.~\cite{falkner_conditioning_2023} showed that such a targeted generation of configurations can be achieved by introducing an additive harmonic bias term with bias center $\hat{\xi}$ and strength $k_{\mathrm{bias}}$ to the potential energy function, which then takes the form
\begin{equation}
    U_{\hat{\xi}}'(x) = U(x) + \frac{k_{\mathrm{bias}}}{2} \left[ \xi(x) - \hat{\xi} \right]^2.
    \label{eq:biased_potential}
\end{equation}
A Boltzmann Generator is then conditioned by including samples from different bias windows in the training set in the case of training by example, appending the respective bias center to the input vector of the multilayer perceptron that parametrizes the bijector. When training by energy, where one samples the prior distribution $p_Z(z)$, each sample $z$ is associated with a condition $c$ drawn from an arbitrary distribution $p(c)$, for example uniformly from a predefined set of conditions. A well-conditioned Boltzmann Generator is then able to sample from arbitrary bias windows even if they were neither represented in the training set nor sampled using $p(c)$.

Leveraging the ability of Boltzmann Generators to draw independent and uncorrelated samples with known importance weights from a biased Boltzmann distribution, one can construct a transition path sampling algorithm~\cite{falkner_conditioning_2023} that alleviates two main limitations of the standard TPS procedure: its inherent sequentiality and the correlations between subsequently sampled paths. The central idea is to use the Boltzmann Generator to produce independent shooting points instead of selecting them from an existing trajectory. To maximize the probability of obtaining a reactive path, those shooting points are generated as close as possible to the transition state region of the system by conditioning the Boltzmann Generator on a pre-defined reaction coordinate. Transition paths are then generated by integrating the equations of motion forward and backward in time starting from the generated shooting points and retaining all paths that connect the two stable states. From the resulting ensemble, the correct transition path distribution can be recovered by re-sampling via the path weights~\cite{falkner_conditioning_2023}
\begin{equation}
    \Omega(X) \propto \left[ \sum_{i = 1}^{L(X)} \frac{\rho_{\mathrm{SP}}^{\hat{\xi}}(x_i)}{\rho(x_i)} \right]^{-1},
    \label{eq:path_weights_general}
\end{equation}
where $X = \{x_1, x_2, \dots, x_{L(X)}\}$ is one such path and $\rho_{\mathrm{SP}}^{\hat{\xi}}(x) \propto e^{-\beta U_{\hat{\xi}}'(x)}$ and $\rho(x) \propto e^{-\beta U(x)}$ are the shooting point and equilibrium Boltzmann distributions, respectively. For this scheme to perform well, it is essential that the chosen reaction coordinate provides an accurate parameterization of the transition process and that the location of the transition state along it is known.

\subsection{The GenAIMMD algorithm}
\label{sec:genaimmd}
\begin{figure}
    \includegraphics[width=8.5cm]{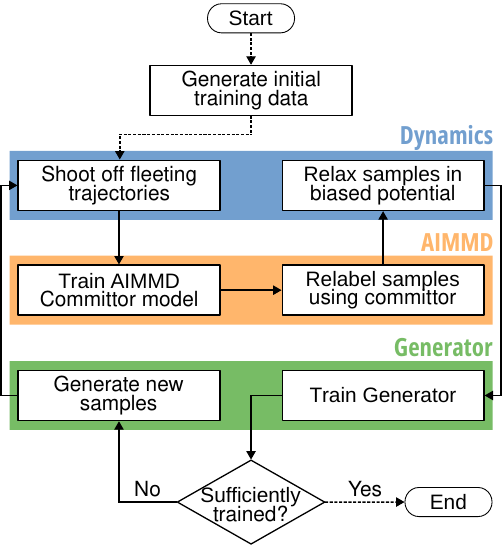}
    \caption{\label{fig:algorithm_flowchart}Overview of the steps involved in the GenAIMMD algorithm. Flowchart elements connected by solid arrows make up the self-consistency loop, while dashed arrows connect steps outside of the loop. The in-loop steps are assigned one of three categories: Dynamics (blue), where the main focus lies on propagating training samples, AIMMD (orange), where the committor model and sample labels are updated, and Generator (green), where the Boltzmann Generator is trained and used to generate new training samples.}
\end{figure}

The central result of this work is GenAIMMD, a multi-step active learning algorithm designed to overcome the difficulty of generating efficient shooting points, and hence uncorrelated transition pathways, when a suitable reaction coordinate is not known a priori. The algorithm uses the learned committor to steer the generation of shooting points towards regions of configuration space with a high chance of generating a transition path (replacing $\xi(x)$ with $p_B(x|\theta)$ in Eq.~\eqref{eq:biased_potential}) and uses these transition paths to refine the estimation of the committor. The active learning loop alternates between three stages: Evolving training samples in a biased or unbiased potential, training the committor model and training the Boltzmann Generator as well as using it to generate new samples. An overview of the algorithm and its three stages is given in Fig.~\ref{fig:algorithm_flowchart}.

\subsubsection*{Self-consistency loop}

To kick-start the training loop, a set of initial training configurations, which we will refer to as \textit{samples} in the following, is required. This set should comprise samples both inside the stable states A and B as well as along transition paths. The size of the dataset depends on the complexity of the system under investigation and can therefore be treated as a hyperparameter, to be tuned by weighing increased early-stage learning performance and stability against the cost of producing additional data.

Given the initial samples, the training loop is repeated until self-consistency is reached. The training paradigm for the AIMMD committor model requires configurations to be labeled with binary shooting outcomes $s_{ij}$. These labels are obtained by shooting off trajectories from the configurations in the training set --- two per configuration --- and recording which of the two states they enter first. The pairs of samples and labels are then used to train the committor model.

Since the Boltzmann Generator's target distribution includes a harmonic bias around given bias centers, each sample used in training the generator must also be associated with such a bias center. We refer to this process as \textit{re-labeling the samples using the committor}, as the bias centers are assigned based on the committor model's prediction for each sample's committor. One can choose to either assign the exact value of the predicted committor to each sample, resulting in each bias window containing exactly one sample, or to bin samples into predefined bias windows based on the prediction. In practice, we found both options to be equally valid and arbitrarily decided to associate each sample with the integer bias center closest to the committor model's prediction in log-space.

This rather arbitrary method of assigning bias centers, however, does not guarantee that the samples are equilibrated in their respective biased potential, making them unsuitable for the \textit{training by example} paradigm of the Boltzmann Generator. Therefore, a short equilibration run must be performed after the re-labeling process. Crucially, this step also ensures that the generator is not trained on the exact same samples it produced at the end of the previous cycle.\cite{shumailov_ai_2024} After the relaxation run is finished, the samples can be used to train the Boltzmann Generator.

Before starting the next cycle, the algorithm checks whether self-consistency has been reached and, if so, the training is terminated. The exit condition must be chosen with care, as an inappropriate choice could either lead to stopping the training prematurely or, conversely, to wasting resources on an already fully trained system. In general, convergence of metrics like the RESS (Eq.~\eqref{eq:ress}) or the loss functions of the Boltzmann Generator and the committor model can serve as reliable predictors of training success.

If the stopping criterion is not met, the loop continues to the next step, where the Boltzmann Generator is used to produce new training samples to be added to the dataset. To limit the growth of the number of samples in the dataset, the oldest members are continuously removed following the \textit{first in, first out} (FIFO) principle. Oftentimes, especially in the early stages of training, a large portion of the generated samples have exceedingly low weight in the target distribution due to the training of the Boltzmann Generator not being complete. Not only are such highly non-physical samples undesirable due to their low importance in training the committor model and Boltzmann Generator, but they also pose a numerical problem when trajectories are shot off from them: in this case, the numerical integration requires the use of much smaller time steps to guarantee stability of the dynamics. To alleviate this, the newly produced samples are first subjected to multiple iterations of re-generation, discarding highly non-physical samples that would break stability and replacing them with newly generated configurations. The procedure is repeated until either a maximum number of iterations is reached or a plateau in the RESS is detected. Each sample generation process is followed by a short equilibration run, where the samples are further relaxed in their respective biased potential. By continuously generating new samples at various committor bias centers, the algorithm is forced to explore the entire target space, adjusting the prediction for the committor when necessary. After relaxation of the new samples, the loop is repeated until self-consistency is reached.

\subsubsection*{Fine-tuning}
\label{sec:fine-tuning}

As discussed in Sec.~\ref{sec:boltzmann_generators}, training a Boltzmann Generator by energy is often not sufficient, since the reverse KL divergence suffers from mode-seeking behavior. This means that, with no target data, full coverage of all the modes of the target distribution is not guaranteed and will not happen in general, particularly when the source and target distributions differ significantly. In previous applications,\cite{wirnsberger_targeted_2020,wirnsberger_normalizing_2021,schebek_efficient_2024,coretti_learning_2025} this issue has often been mitigated by choosing source and target distributions that are sufficiently similar, which, in some cases, allows training by energy alone. In our case, however, the two distributions are considerably different. We therefore supply a few samples from the target distribution to improve mode coverage through maximum likelihood estimation (i.e., training by example). However, if the samples provided do not sufficiently cover the main modes of $p_X(x)$, the generator may still drastically under-sample regions that carry significant probability mass in the target distribution. The few samples produced in such regions will then have exceedingly large importance weights, which can destabilize the re-weighting procedure. Although the presented algorithm automatically generates additional training samples, these are unlikely to land in yet unexplored modes that were not represented in the initial training set.

To address this issue, one can encourage the generator to explore the under-sampled modes by deliberately training it on samples from these regions. We apply this fine-tuning to the Boltzmann Generator as it leaves the GenAIMMD loop by
\begin{enumerate}
    \item \label{fine-tune:start} generating a large set of samples from the learned distribution in the bias window associated with the transition state characterized by $\hat{p}_B = 0.5$,
    \item computing the associated importance weights,
    \item smoothing the high-variance importance weights via Pareto Smoothed Importance Sampling (PSIS),\cite{vehtari_pareto_2024}
    \item re-sampling the generated samples using the smoothed weights into the biased Boltzmann distribution,
    \item \label{fine-tune:end} training the Boltzmann Generator by example using the re-sampled, fully synthetic samples and
    \item repeating steps \ref{fine-tune:start} -- \ref{fine-tune:end} until an appropriate convergence criterion (e.g. a plateau of the ESS) is met.
\end{enumerate}
The reason we restrict the sampling and training to the $\hat{p}_\mathrm{B} = 0.5$ window is that, after training, the Boltzmann Generator is only used to produce shooting points inside this bias window. In principle, however, this procedure can be expanded to an arbitrary number of different bias windows. Note that training on synthetic data is not problematic here,\cite{shumailov_ai_2024} since we previously did all the training on real samples and merely seek to reduce the variance of the importance weights. Moreover, generated samples are always re-weighted or re-sampled into the correct Boltzmann distribution.

\subsubsection*{Obtaining transition paths}

Once the Boltzmann Generator has been trained and, if necessary, fine-tuned, shooting points can be generated in the transition state region defined by $p_{\mathrm{B}} \approx 0.5$. From this generated set of shooting points, trajectories can be shot off and connected to form transition paths. Inserting Eq.~\eqref{eq:biased_potential} into Eq.~\eqref{eq:path_weights_general} with the learned committor $p_{\mathrm{B}}(x | \theta)$ as the CV bias yields the expression
\begin{equation}
    \Omega(X) \propto \left[ \sum_{i=1}^{L(X)} \exp \left( -\beta \frac{k_{\mathrm{bias}}}{2} \left[ p_{\mathrm{B}}(x_i | \theta) - \hat{p}_{\mathrm{B}} \right]^2 \right) \right]^{-1}
\end{equation}
for the corresponding path weights, where $\hat{p}_{\mathrm{B}} = 0.5$ is the committor bias center.
\section{Results}
\label{sec:results}

\subsection{Two-dimensional model}

\begin{figure}
    \includegraphics[width=8.5cm]{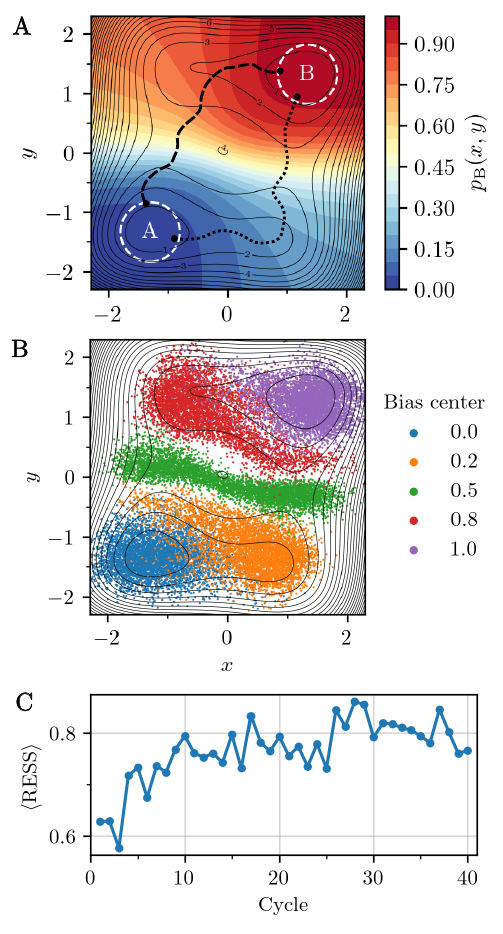}
    \caption{\label{fig:wolfe_quapp}Training results for the two-dimensional Wolfe-Quapp potential. (A) shows the iso-lines of the potential energy in solid black with labels in units of $k_{\mathrm{B}}T$, as well as the committor $p_{\mathrm{B}}(x, y)$ corresponding to the states defined by the white circles, learned using the GenAIMMD algorithm. The two black dashed lines connecting the states illustrate transition paths using the upper (long dashes) and the lower (short dashes) reaction channel, respectively. (B) depicts samples generated by the Boltzmann Generator in different committor bias windows and re-sampled into the corresponding biased Boltzmann distribution (see Eq.~\eqref{eq:biased_potential}). (C) plots the mean relative effective sample size (Eq.~\eqref{eq:ress}) as a function of algorithm cycles, where the average is taken across multiple committor bias windows.}
\end{figure}

To assess the efficacy of GenAIMMD, a two-dimensional system with two stable states is considered first, namely a rotated, shifted and scaled form of the Wolfe-Quapp potential~\cite{wolfe_chemical_1975, quapp_growing_2005} depicted in Fig.~\ref{fig:wolfe_quapp}A. The set of initial samples passed to the GenAIMMD algorithm consists of 100 samples inside each of the two stable states and 100 samples along transition paths obtained through TPS, summing up to a total of 300 initial samples. In this work, we restrict the dynamics to the overdamped regime in which velocities can be neglected.

The GenAIMMD loop is run for a total of 50 cycles, and after each cycle, the current RESS (Eq.~\eqref{eq:ress}) is recorded and used to measure convergence (see Fig.~\ref{fig:wolfe_quapp}C). After approximately 10 cycles (107 seconds on an RTX 5070 GPU), the RESS plateaus at $\approx 0.8$, indicating that self-consistency has been reached. The Boltzmann Generator is then able to sample from the biased Boltzmann distribution arising from Eq.~\eqref{eq:biased_potential} with the committor model used as a reaction coordinate at arbitrary committor bias centers, as can be seen in Fig.~\ref{fig:wolfe_quapp}B. For this system, no sign of poor mode coverage was observed during reweighting and therefore no fine tuning was performed after the GenAIMMD  training loop.

\subsection{Polymer model}

\begin{figure*}
    \includegraphics[width=17cm]{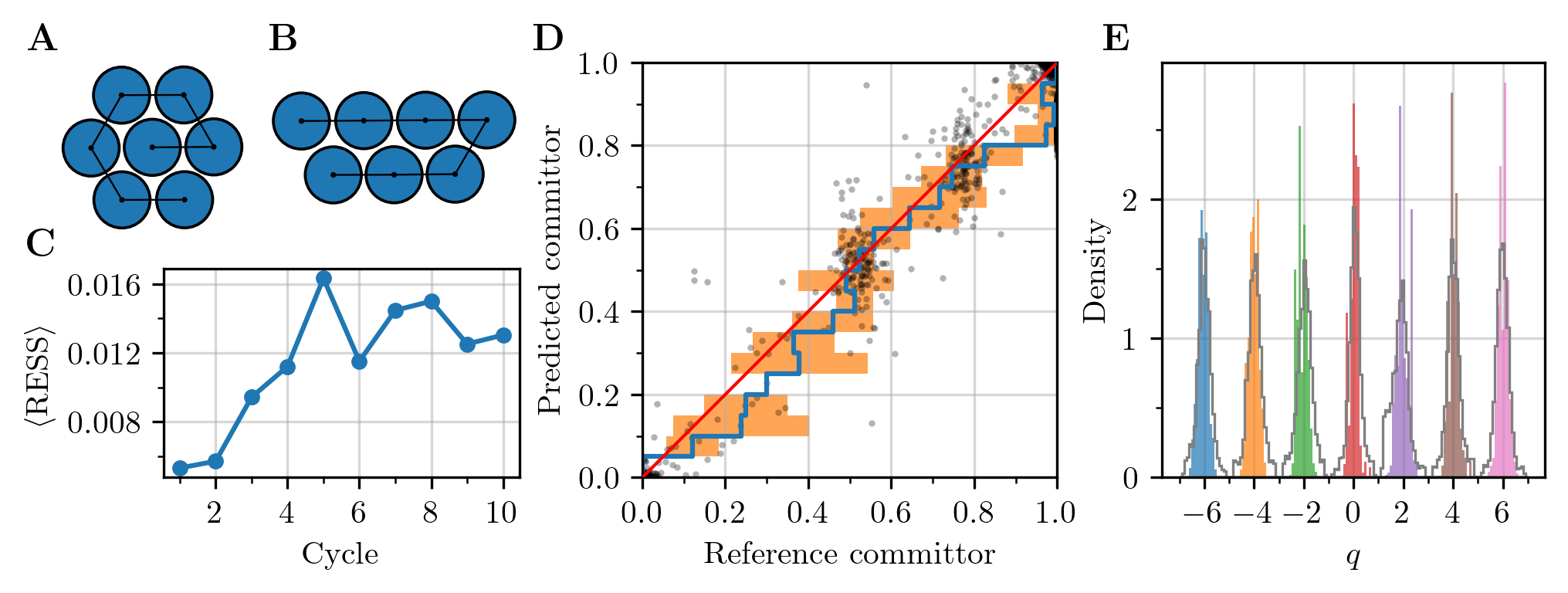}
    \caption{\label{fig:polymer}Training results for the polymer model. (A) and (B) depict the stable conformations of the polymer, i.e., state A and state B, respectively. (C) shows the relative effective sample size as a function of GenAIMMD cycles, averaged over multiple committor bias windows within each cycle. (D) displays the predicted committor against the numerically estimated (reference) committor for a set of test samples, where each black dot corresponds to a single system configuration and the straight red line corresponds to perfect agreement between the reference and the prediction. The blue line marks the average reference committor binned by the predicted committor, and the orange area indicates the corresponding standard deviation. (E) histogram of the log-committor $q(x)$ (Eq.~\eqref{eq:p_of_q}) of polymer samples generated using the Boltzmann Generator at seven different bias centers (the labels of the $q$-axis are the bias centers in log-committor space), along with reference histograms (gray outlines) obtained from Umbrella Sampling. Note that $q \to -\infty \Leftrightarrow p_{\mathrm{B}} \to 0$, $q \to +\infty \Leftrightarrow p_{\mathrm{B}} \to 1$ and $q =0 \Leftrightarrow p_{\mathrm{B}} = 0.5$.}
\end{figure*}

Following the simulations in two dimensions, we increase the learning difficulty by moving to a higher-dimensional system --- again using overdamped Langevin dynamics. We consider a linear polymer in two spatial dimensions that consists of seven monomers, resulting in a total of 11 degrees of freedom after subtracting the center of mass and rotation of the polymer.\cite{falkner_conditioning_2023} The non-bonded interactions between the monomers are modeled through a conventional 12-6 Lennard-Jones potential, which is complemented by two bonded contributions: harmonic bond stretching and a cosine-based angle potential. This system has multiple (meta)stable states, from which we choose a curled-up conformation (Fig.~\ref{fig:polymer}A) and an extended conformation (Fig.~\ref{fig:polymer}B) for the purpose of testing our method. In our study, the radius of gyration is used as order parameter to distinguish the two stable states and the transition region.

To account for the translational and rotational invariance of the system, we introduce an invertible internal coordinate map that transforms Cartesian coordinates to normalized bond lengths and bond angles. To make the system invariant with respect to reflections, we constrain the first bond angle $\varphi_1$ to the upper half of the unit circle, mirroring the polymer along the $x$-axis if $\sin \varphi_1 < 0$. To improve training stability, we condition the flow on the log-committor $q(x|\theta)$ (see Eq.~\eqref{eq:p_of_q}) instead of the committor $p_{\mathrm{B}}(x|\theta)$.

The mean RESS (Fig.~\ref{fig:polymer}C) observed while running the GenAIMMD algorithm for this system is significantly lower compared to the two-dimensional model, which is expected due to the higher dimensionality of the learning problem. Convergence of the RESS was reached after about five GenAIMMD cycles.

To provide a reference that can be used to assess the accuracy of the trained committor model, we first generate polymer samples along an artificial reaction coordinate (in this case the radius of gyration) using Umbrella Sampling.\cite{torrie_nonphysical_1977, kastner_umbrella_2011} From each sample, we initiate 1000 fleeting trajectories and observe the states they hit first, which yields a numerical estimate for their committor values. Then, we plot these reference committor values against the committor model's predictions for the samples (Fig.~\ref{fig:polymer}D). The close clustering of the predictions along the $y = x$ line shows a very good agreement between the committor model and the reference.

Lastly, we verify that the Boltzmann Generator is able to sample from the biased Boltzmann distribution. For a set of arbitrarily chosen committor bias centers that span a broad range of possible committor values, including that which corresponds to the transition state ($\hat{p}_{\mathrm{B}} = 0.5$), we produce polymer samples using the Boltzmann Generator and histogram their predicted log-committor, re-weighting each sample with the corresponding importance weight (Eq.~\eqref{eq:boltzmann_weights}). As reference, we use the learned committor as a reaction coordinate and perform Umbrella Sampling using the same bias windows. The resulting histogram is shown in Fig~\ref{fig:polymer}E. The spikes visible in the reweighted distribution are a sign of disproportionately high importance weights of specific configurations, which are related to the problem discussed in Sec.~\ref{sec:genaimmd} and in previous publications.\cite{midgley_flow_2023} Therefore, before generating shooting points to be used in path sampling in the next section, we apply the fine-tuning procedure described in Sec.~\ref{sec:genaimmd} to the Boltzmann Generator in the bias window corresponding to $\hat{q} = 0$.

\subsection{Application to Transition Path Sampling}

\begin{figure*}
    \includegraphics[width=17cm]{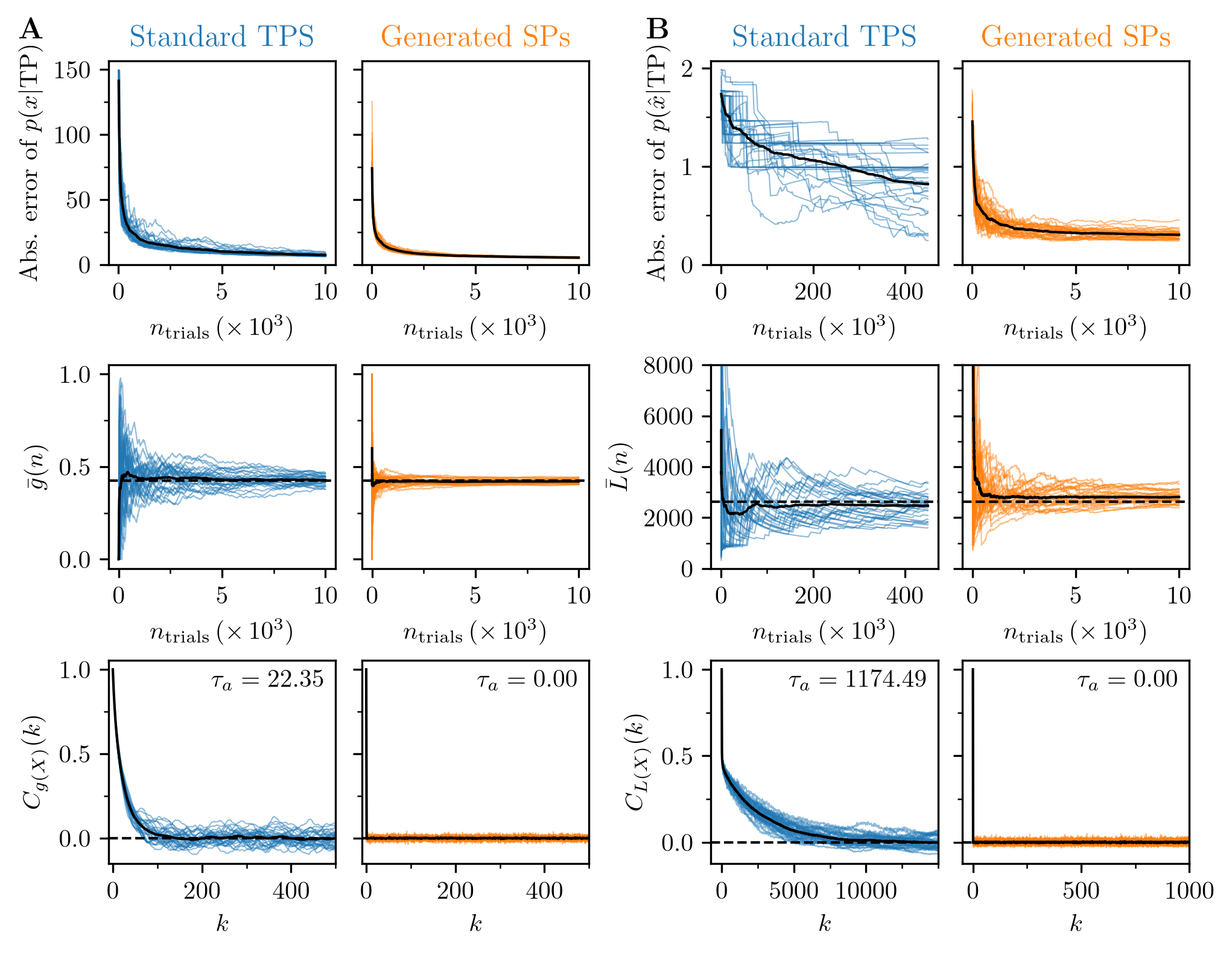}
    \caption{\label{fig:tps_comparison}Performance comparison between standard TPS (blue) and shooting points generation using the GenAIMMD-trained Boltzmann Generator conditioned on the learned committor (orange) in (A) the two-dimensional model and (B) the polymer model. The colored lines are 30 independent runs (TPS simulations in case of standard TPS, blocks of generated shooting points in case of the sampled shooting points). The solid black lines are averages over the colored lines at the respective step. First row: Absolute error of the distribution of points along transition paths $p(x|\mathrm{TP})$ as a function of two-way shooting trials, measured as the $L^1$~distance between a reference histogram obtained via a long-running TPS simulation and the histogram arising from the respective method after $n_{\mathrm{trials}}$ trials (see the Supplementary Material for more information). Second row: Running average of the reaction channel indicator function $g(X)$ (A) and the path length $L(X)$ (B) after $n_{\mathrm{trials}}$ two-way shooting trials. Dashed lines are averages from the reference simulation. Third row: Autocorrelation function $C_{F(X)}(k)$ and integrated autocorrelation time $\tau_a$ of $F(X)$, which is the function averaged in the second row of (A) and (B), respectively.}
\end{figure*}

To illustrate the advantages of independently drawing highly reactive shooting points, we benchmark GenAIMMD against conventional two-way shooting TPS for both of the presented test systems. The results are shown in Fig.~\ref{fig:tps_comparison}A for the two-dimensional model and Fig.~\ref{fig:tps_comparison}B for the polymer model. As reference, we use long-running uniform-shooting-point-selection TPS simulations in both systems (see the Supplementary Material for details). For the performance analysis, we use 30 replicas of the same simulation for both methods whose resulting observables are plotted as thin colored lines in Fig.~\ref{fig:tps_comparison}. The thick black lines represent the average over these replicas.

Histogramming the individual configurations along transition paths, one obtains the distribution $p(x | \mathrm{TP)}$. For the polymer model, we divide the configurations into discrete classes $\hat{x}$ based on their bond angles, which then serve as histogram bins. Such histograms are created for both test systems (Wolfe-Quapp potential and polymer) and both path sampling methods (standard TPS and Generated Shooting Points). We then proceed to calculate the $L^1$ distance
\begin{equation}
    d(p, q) = \sum_{i = 1}^N |p_i - q_i|
\end{equation}
between the bins $p = (p_1, p_2, \dots, p_N)$ of these histograms and those of the respective reference histogram $q = (q_1, q_2, \dots, q_N)$ as a function of the number of two-way shooting trials $n_{\mathrm{trials}}$, which we report as the absolute error of $p(x|\mathrm{TP)}$ in the first row of Fig.~\ref{fig:tps_comparison}. In the case of generated SPs, a trial is considered the generation of a shooting point and the integration of two trajectories, which are then stitched together to form a path, as described in Sec.~\ref{sec:tps}. Trials that produce non-reactive paths are still counted towards the total number of trials; however, the corresponding paths are assigned zero weight.

Standard TPS is known to suffer from correlations between subsequently sampled paths, which often leads to trapping inside one reaction channel for an extended period of time. The two-dimensional model has two such reaction channels that transition paths may use (black dashed lines in Fig.~\ref{fig:wolfe_quapp}A), and we distinguish them via the indicator function $g(X)$, which is 1 if a path $X$ uses the upper channel and 0 if it uses the lower one. Computing the average $\bar{g}(n) = \frac{1}{n} \sum_{i=1}^n g(X_i)$ of $g(X)$ after $n$ shooting attempts, one converges to the relative population of the two channels as $n \to \infty$, which depends on their free energy difference and is indicated by the dashed line in the second row of Fig.~\ref{fig:tps_comparison}A. The plots show the convergence behavior of this average for standard TPS and the shooting points originating from the Boltzmann Generator. The third row of Fig.~\ref{fig:tps_comparison}A then depicts the normalized autocorrelation function $C_{g(X)}(k)$ and integrated autocorrelation time $\tau_a$ of $g(X)$ for the two methods.

In the polymer model, a function that distinguishes the reaction channels is not readily available, which is why we instead choose to investigate the convergence of the mean path length $\bar{L}(n) = \frac{1}{n} \sum_{i = 1}^n L(X_i)$ after $n$ shooting trials. This is shown in the second row of Fig.~\ref{fig:tps_comparison}B along with the corresponding autocorrelation function $C_{L(X)}(k)$ and integrated autocorrelation time $\tau_a$ in the third row.
\section{Discussion and conclusion}
\label{sec:discussion}
In this work, we introduced a new method we call GenAIMMD that enables efficient shooting point generation for TPS using generative machine learning~\cite{falkner_conditioning_2023} without requiring a priori knowledge of a suitable reaction coordinate. Since the shooting points obtained in this way are fully independent of each other, the resulting transition paths are free from correlations and can be harvested in parallel, overcoming two shortcomings of the established shooting algorithm. The method also yields the committor for the process. At its core, the method employs an active learning loop that alternates between committor learning through AIMMD~\cite{jung_machine-guided_2023} and training a Boltzmann Generator conditioned on the learned committor. The generator then seeds new training samples, and the loop continues until self-consistency is achieved. In comparison to recently proposed methods,\cite{tang_breaking_2026} the exact-likelihood nature of Boltzmann Generators enables us to re-weight generated configurations into the correct distribution, allowing computation of unbiased thermodynamic estimates in addition to exploring the transition region.

We tested our method on two systems: the two-dimensional Wolfe-Quapp potential and a polymer model with eleven degrees of freedom. In both systems, GenAIMMD succeeded in training a Boltzmann Generator as well as learning the committor. Employing the same shooting point resampling and path re-weighting scheme as in Ref.~\onlinecite{falkner_conditioning_2023}, we demonstrated that GenAIMMD manages to outperform standard TPS by orders of magnitude while maintaining independence of a predetermined reaction coordinate.

Nevertheless, the method has some limitations. As mentioned above, the success of GenAIMMD at covering all the target distribution's modes depends strongly on their representation in the initial training set because the KL divergence used to train by energy promotes mode-seeking behavior. These effects were already evident in the polymer model, where they were successfully mitigated by fine-tuning the Boltzmann Generator before producing shooting points. However, we expect them to intensify with growing system size and number of modes. As a possible solution, other training objectives, such as $\alpha$-divergence, have been proposed~\cite{midgley_flow_2023} because they are mass-covering and minimize the variance of the importance weights. Another limiting factor is the well-recognized decline in sampling efficiency of Boltzmann Generators with increasing system dimensionality,\cite{schebek_scalable_2026} which is due to the exponential accumulation of error between the learned and target distributions. Moreover, our current implementation uses affine transformations in the Boltzmann Generator's coupling layers, which not only limits the model's expressivity but also prevents us from appropriately modeling periodic coordinates. However, the GenAIMMD scheme is largely agnostic to the architecture used to parameterize the diffeomorphism in the Boltzmann Generator. This makes it straightforward to incorporate more expressive transformations, such as those proposed in Refs.~\citenum{durkan_neural_2019,rezende_normalizing_2020,kohler_smooth_2021}, in place of the affine transformations used here. Future implementations will further explore architectures specifically designed to improve scalability to higher-dimensional systems~\cite{tan_scalable_nodate,schebek_scalable_2026} and to appropriately handle periodic coordinates,\cite{wirnsberger_normalizing_2021} while also investigating more expressive transformations and strategies to mitigate mode collapse. We are confident that these developments will allow GenAIMMD to be extended to increasingly complex systems while addressing the limitations associated with the dimensionality and representational capacity of the current implementation.
\section*{Supplementary Material}

The supplementary material contains additional information about the presented test systems (system definition, internal coordinate transformation), training and simulation parameters, reference data and details about the benchmarking against standard TPS.

\section*{Acknowledgments}

This paper is dedicated to Gerhard Hummer on the occasion of his 60th birthday. We are grateful for many years of stimulating discussions and scientific exchange. Gerhard has been a role model and an inspiration for us and the entire community, and we look forward to many more years of friendship, collaboration, and shared scientific adventures.

This research was funded in part by the Austrian Science Fund (FWF) through SFB TACO 10.55776/F8100 and in part by 10.55776/COE5 (Cluster of Excellence MECS), both of them available via https://www.fwf.ac.at/en/discover/research-radar. For open access purposes, the author applied a CC BY public copyright license to any author-accepted manuscript version arising from this submission.

The computational results presented have been achieved in part using the Vienna Scientific Cluster (VSC).

\section*{Author Declarations}

\subsection*{Conflict of Interest}

The authors have no conflicts to disclose.

\subsection*{Author Contributions}

\textbf{Maximilian Negedly}: Conceptualization (equal); Formal Analysis (equal); Investigation (lead); Methodology (equal); Software (lead); Visualization (lead); Writing --- original draft (lead).

\textbf{Sebastian Falkner}: Conceptualization (equal); Formal Analysis (equal); Investigation (supporting); Methodology (equal); Supervision (equal).

\textbf{Alessandro Coretti}: Formal Analysis (equal); Investigation (supporting); Methodology (equal); Supervision (equal); Writing --- original draft (supporting).

\textbf{Christoph Dellago}: Conceptualization (equal); Formal Analysis (supporting); Funding Acquisition (lead); Methodology (equal); Project Administration (lead); Supervision (equal); Writing --- original draft (supporting).

\section*{Data availability}

The data that support the findings of this study are openly available on GitHub at https://github.com/CompPhysVienna/paper\_genaimmd.

\onecolumngrid
\clearpage
\normalsize

\setcounter{figure}{0}
\setcounter{table}{0}
\setcounter{equation}{0}
\setcounter{section}{0}

\setlength{\parskip}{0.3cm}
\setlength{\parindent}{0cm}

\renewcommand{\thesection}{S\arabic{section}}
\renewcommand{\thesubsection}{\thesection.\arabic{subsection}}
\renewcommand{\thesubsubsection}{\thesubsection.\arabic{subsubsection}}

\renewcommand{\theequation}{S\arabic{equation}}
\renewcommand{\thefigure}{S\arabic{figure}}
\renewcommand{\thetable}{S\arabic{table}}

\renewcommand{\figurename}{Supplementary Figure}
\renewcommand{\tablename}{Supplementary Table}

\renewcommand{\theHsection}{S\arabic{section}}
\renewcommand{\theHsubsection}{S\arabic{section}.\arabic{subsection}}
\renewcommand{\theHsubsubsection}{S\arabic{section}.\arabic{subsection}.\arabic{subsubsection}}
\renewcommand{\theHequation}{S\arabic{equation}}
\renewcommand{\theHfigure}{S\arabic{figure}}
\renewcommand{\theHtable}{S\arabic{table}}

{\Large\textbf{Supplementary material for ``Correlation-Free Transition Path Sampling through Shooting Point Generation Guided by Committor Learning''}}

\begin{center}
    Maximilian Negedly\textsuperscript{1, 2}, Sebastian Falkner\textsuperscript{1, 3}, Alessandro Coretti\textsuperscript{1} and Christoph Dellago\textsuperscript{1, a)}

    \textit{\textsuperscript{1}Faculty of Physics, University of Vienna, 1090 Vienna, Austria} \\
    \textit{\textsuperscript{2}Vienna Doctoral School in Physics, University of Vienna, 1090 Vienna, Austria} \\
    \textit{\textsuperscript{3}Institute of Physics, University of Augsburg, 86159 Augsburg, Germany}

    a) Electronic mail: christoph.dellago@univie.ac.at
\end{center}
\section{Two-dimensional model}

This section contains general information about the two-dimensional system used to test the GenAIMMD algorithm, as well as reference data for some of the figures in the main text and the hyperparameters used in training.

\subsection{System definition}
\begin{figure}[h]
    \centering
    \includegraphics[width=\textwidth]{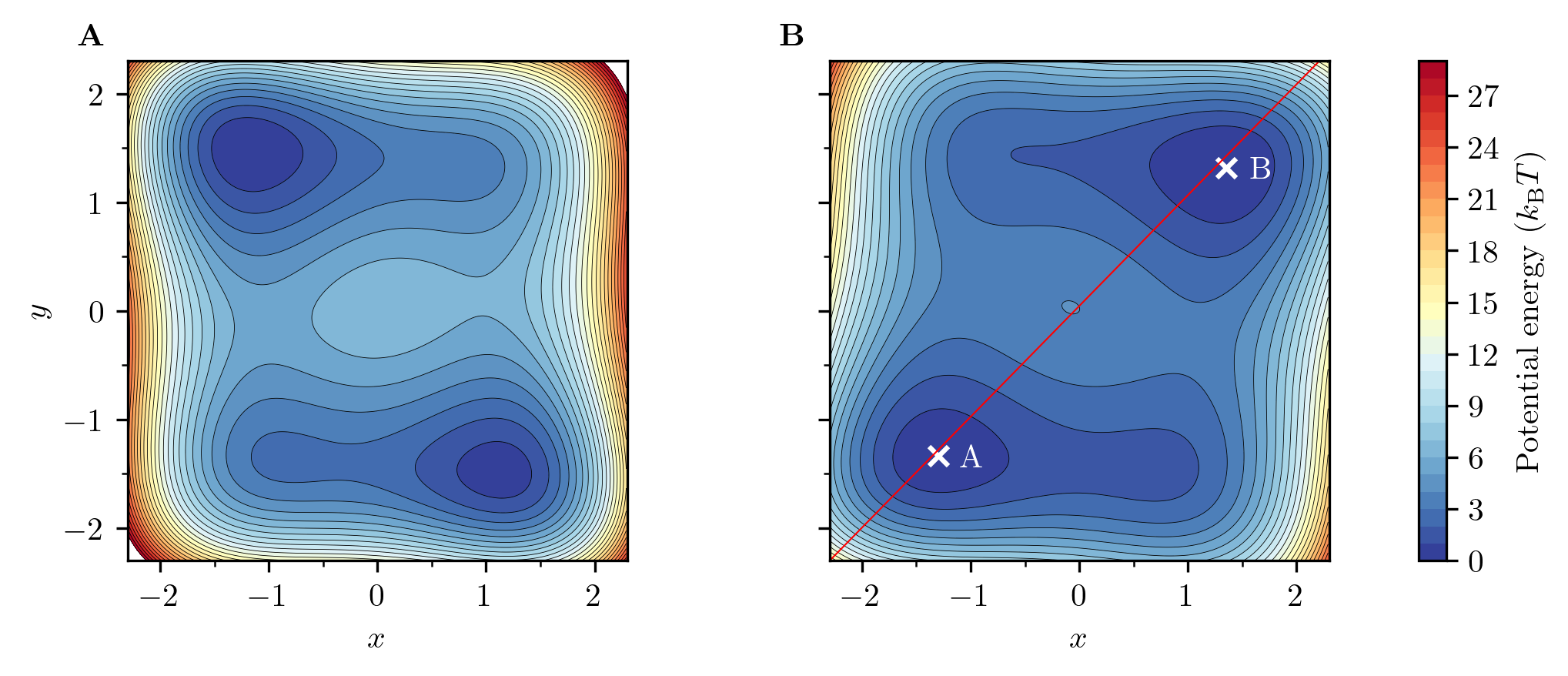}
    \caption{The two-dimensional system definition and transformation. (A) shows the original form of the Wolfe-Quapp potential~\cite{wolfe_chemical_1975_supp, quapp_growing_2005_supp}. (B) depicts the rotated, shifted and scaled form used in this work, which aligns the centers of states A and B with the red line defined by $y = x$.}
    \label{supp_fig:wolfe_quapp_definition}
\end{figure}

The functional form of the two-dimensional model is based on the Wolfe-Quapp potential energy function~\cite{wolfe_chemical_1975_supp, quapp_growing_2005_supp}
\begin{equation}
    U_{\text{WQ}}(x, y) = x^4 + y^4 - 2x^2 - 4y^2 + xy + 0.3x + 0.1y
    \label{supp_eq:wq_unnormalized}
\end{equation}
depicted in Supplementary Figure~\ref{supp_fig:wolfe_quapp_definition}A. We choose to rotate this potential energy function such that its minima --- the center points of the circles with radii $r = 0.5$ defining states A and B (Fig.~2A in the main text) --- lie on the artificially constructed reaction coordinate defined by the line $y(x) = x$, which we later use to distinguish the two reaction channels of the system. In addition to rotating the potential, we make the barrier height $h$ variable by multiplying Eq.~\eqref{supp_eq:wq_unnormalized} by the appropriate normalization factor, yielding
\begin{equation}
    \begin{split}
        U_{\text{WQ}}'(x, y) = \frac{h}{c_{11}} \cdot \Big[ &c_1 \cdot x^4 + c_2 \cdot x^3 y + c_3 \cdot x^2+ c_4 \cdot x^2 y^2 + c_5 \cdot x \\
        &+ c_6 \cdot x y+ c_7 \cdot x y^3+ c_8 \cdot y + c_9 \cdot y^2+ c_{10} \cdot y^4 + c_{11}\Big]
    \end{split}
    \label{supp_eq:wq}
\end{equation}
with coefficients $c_1 = 0.969233$, $c_2 = -0.480614$, $c_3 = -2.155285$, $c_4 = 0.184602$, $c_5 = -0.285146$, $c_6 = -1.46487$, $c_7 = 0.480614$, $c_8 = 0.136719$, $c_9 = -3.84471$ and $c_{10} = 0.969233$, as well as $c_{11} = 6.76245$, which shifts the global minimum to zero. In this work, we set $h = 4 \, k_{\mathrm{B}}T$. The resulting potential energy function (Eq.~\eqref{supp_eq:wq}) is shown in Supplementary Figure~\ref{supp_fig:wolfe_quapp_definition}B.

\subsection{Reference data}

\begin{figure}[h]
    \centering
    \includegraphics[width=\textwidth]{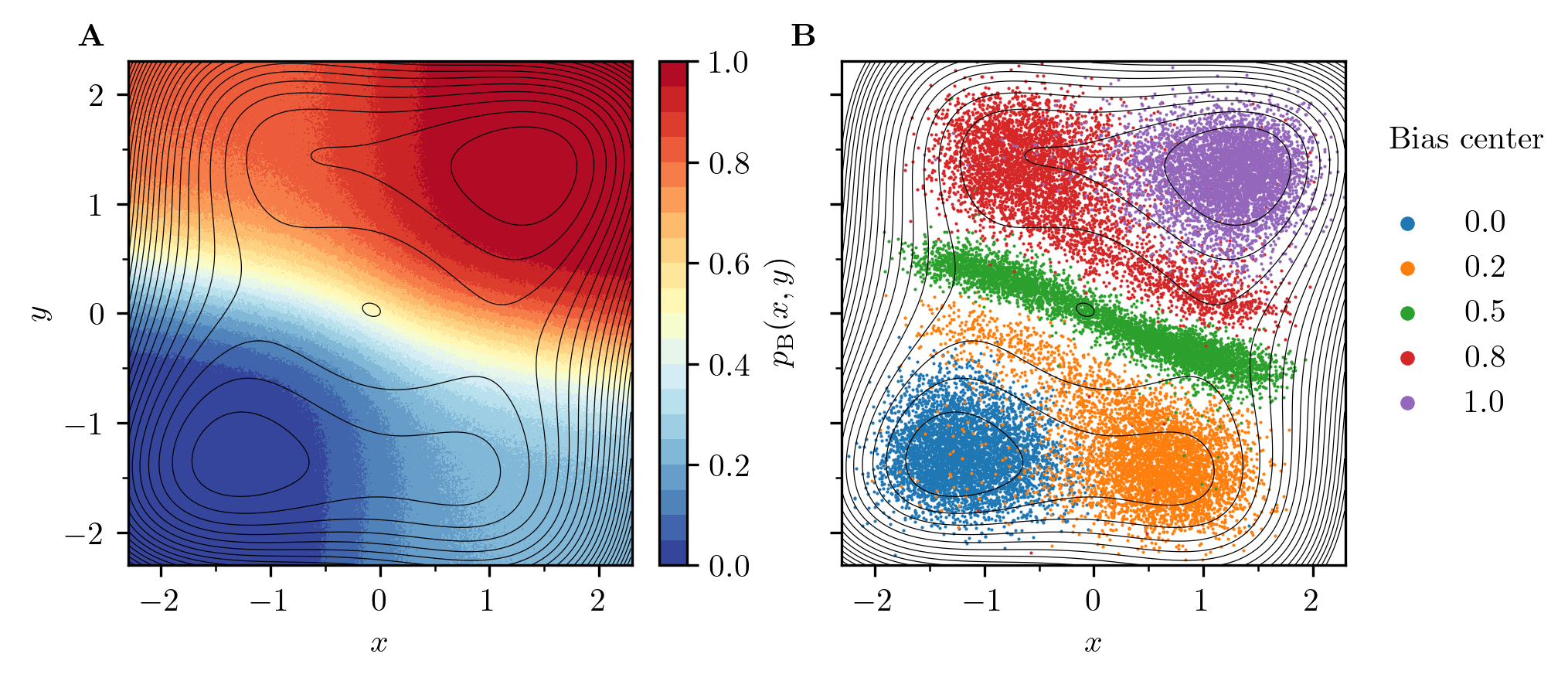}
    \caption{Two-dimensional system reference plots. (A) contains the reference committor $p_{\mathrm{B}}(x, y)$ on an evenly spaced grid on $x$ and $y$, obtained by initiating 1000 fleeting trajectories from each point on the grid and observing which state they enter first. (B) shows system configurations equilibrated in the Wolfe-Quapp potential biased by the reference committor at different committor bias centers listed on the right.}
    \label{supp_fig:wolfe_quapp_reference}
\end{figure}

To obtain a reference for the committor $p_{\mathrm{B}}(x, y)$ in the two-dimensional system, we initiate 1000 fleeting trajectories from points on a uniform grid in $x$ and $y$ spaced by $a = 0.01$ in each direction in the range $-2.3 \leq x, y \leq 2.3$. We then record the number of trajectories per lattice point that enter state B before state A and divide this by the total number of trajectories to obtain an estimate for the committor at that point. The results are depicted in Supplementary Figure~\ref{supp_fig:wolfe_quapp_reference}A.

Linearly interpolating this numerically estimated committor between grid points, we construct the harmonic bias potential (Eq.~(10) in the main text) and use it to obtain equilibrium samples from various committor bias windows. These serve as reference for the samples produced by the Boltzmann Generator in the same bias windows (Fig.~2B in the main text). The equilibrated samples are depicted in Supplementary Figure~\ref{supp_fig:wolfe_quapp_reference}B.

\subsection{Training parameters}

\begin{table}
    \centering
    \caption{Parameters used for testing GenAIMMD in the two-dimensional system.}
    \begin{tabular}{lccc}
        \toprule
        & \textbf{Boltzmann Generator} & \textbf{Committor model} & \textbf{GenAIMMD}\\
        \midrule
        Hidden layers & 3 & 2 & \\
        Nodes per layer & 100 & 20, 10 & \\
        Learning rate & $10^{-4}$ & $10^{-4}$ & \\
        Batch size & 100 & 100 & \\
        Epochs per cycle & 200 & 200 & \\
        RealNVP blocks & 3 &  & \\
        $\lambda_{\text{KL}}$ & 1 &  & \\
        $\lambda_{\text{ML}}$ & 1 &  & \\
        \midrule
        Cycles & & & 50 \\
        Diffusion coefficient & & & 1.0 \\
        Timestep & & & $10^{-2}$ \\
        Relaxation MD steps & & & 500 \\
        $k_{\text{bias}}$ & & & 1500 \\
        FIFO queue size & & & $2 \cdot N_{\text{initial samples}}$ \\
        $N_{\text{initial samples}}$ & & & 300 \\
        \bottomrule
    \end{tabular}
    \label{supp_tab:params_wq}
\end{table}

The hyperparameters used to define and train the Boltzmann Generator and the committor model as well as to run the GenAIMMD algorithm are listed in Supplementary Table~\ref{supp_tab:params_wq}. We use overdamped Langevin (Brownian) dynamics integrated via the Euler-Maruyama method, the parameters of which are listed in the same table.

\section{Polymer model}

This section details how the polymer model is defined in terms of the contributions to the potential energy function and the internal coordinate transformation. We also list the hyperparameters used for running GenAIMMD.

\subsection{System definition}

The potential energy function of the polymer model~\cite{falkner_conditioning_2023_supp} has three contributions: a Lennard-Jones term
\begin{equation}
    U_{\mathrm{LJ}} = 4 \varepsilon \sum_{i = 1}^N \sum_{j > i}^N \left[ \left( \frac{\sigma}{r_{ij}} \right)^{12} - \left( \frac{\sigma}{r_{ij}} \right)^6 \right]
\end{equation}
with distance $r_{ij} = |\vec{x}_i - \vec{x}_j|$ between monomers $i$ and $j$ modeling the non-bonded interactions, a harmonic bond-stretching potential
\begin{equation}
    U_{\text{bond}} = \frac{k_{\text{bond}}}{2} \sum_{i = 1}^{N - 1} (r_{i, i+1} - r_{\text{ref}})^2
\end{equation}
for the $N -1$ bonds with preferred length $r_{\text{ref}} = \sqrt[6]{2} \sigma$ between monomers and a cosine-based angle potential
\begin{equation}
    U_{\text{angle}} = \frac{k_{\text{angle}}}{2} \sum_{i = 1}^{N - 2} \left[ 1 - \cos(\varphi_{i, i+1, i+2} - \varphi_{\text{ref}}) \right]
\end{equation}
for the $N - 2$ angles $\varphi_{i, i+1, i+2}$ between three consecutive monomers in the chain, where $\varphi_{\text{ref}} = \pi \, \, \text{rad}$. For the force constants of the two bonded terms, we use $k_{\text{bond}} = 5 \, \varepsilon \sigma^{-2}$ and $k_{\text{angle}} = 1.4 \, \varepsilon$, respectively. The three contributions then combine to form the total potential energy
\begin{equation}
    U_{\text{polymer}} = U_{\text{LJ}} + U_{\text{bond}} + U_{\text{angle}}
\end{equation}
of the system.

The stable states of the polymer model are defined using the radius of gyration
\begin{equation}
    R_{\mathrm{G}}(x) = \sqrt{\frac{1}{N} \sum_{i = 1}^N |\vec{x}_i - \langle x \rangle|^2},
\end{equation}
where $|\vec{x}_i - \langle x \rangle|$ is the distance between the $i$-th monomer $\vec{x}_i$ and the polymer's center of mass $\langle x \rangle$. State A is then defined as $R_{\mathrm{G}} < 1.05 \, \sigma$ and state B as $R_{\mathrm{G}} > 1.2 \, \sigma$.

\subsection{Internal coordinates}

\begin{figure}
    \centering
    \includegraphics[width=\textwidth]{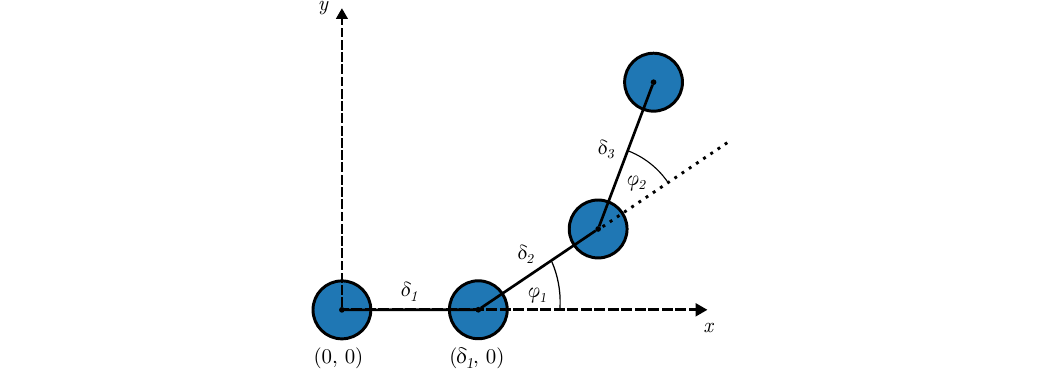}
    \caption{The internal coordinates used in the polymer model~\cite{falkner_conditioning_2023_supp} to account for the system's translational and rotational invariance. These coordinates include bond lengths $\delta_i$ and bond angles $\varphi_i$. The first and second monomers are placed on the x-axis when transforming back to Cartesian coordinates, with the first monomer being placed at the origin of the coordinate system.}
    \label{supp_fig:polymer_internal_coords}
\end{figure}

As described in the main text, to account for the system's invariance under translation and rotation, we transform system configurations from Cartesian to internal coordinates~\cite{falkner_conditioning_2023_supp} before passing them to the Boltzmann Generator or the committor model. These internal coordinates are the $N-1$ bond lengths and the $N-2$ angles between three consecutive monomers --- a total of $2N - 3 = 11$ degrees of freedom for the 7 monomers considered in this work. Internally, these coordinates are represented as vectors of the form $(\delta_1, \delta_2, \dots, \delta_{N - 1}, \varphi_1, \varphi_2, \dots, \varphi_{N - 2})^{\text{T}}$ with $\delta_i := r_{i, i+1}$ and $\varphi_i := \varphi_{i, i+1, i+2}$ (see Supplementary Figure~\ref{supp_fig:polymer_internal_coords}). When transforming back to Cartesian coordinates, we fix the position of the first monomer to $(x_1, y_1) = (0, 0)$ and that of the second monomer to $(x_2, y_2) = (\delta_1, 0)$. The determinant of the resulting Jacobian is
\begin{equation}
    \det J = \prod_{i = 2}^{N-1} \delta_i,
\end{equation}
and we refer to Sec.~SVII of the Supplementary Information of Ref.~\onlinecite{falkner_conditioning_2023_supp} for a full derivation.

\subsection{Normalization layer}

To assist with training, we normalize the internal coordinates by subtracting the mean $\mu_i$ and dividing by the standard deviation $\sigma_i$ of each coordinate computed from the initial training set. So, for each internal coordinate $z_i \in (\delta_1, \delta_2, \dots, \delta_{N - 1}, \varphi_1, \varphi_2, \dots, \varphi_{N - 2})$ with $i = 1, \dots, 2N-3$,
\begin{equation}
    \hat{z}_i = \frac{z_i - \mu_i}{\sigma_i}.
\end{equation}
The resulting log Jacobian determinant is
\begin{equation}
    \log \det J = \sum_{i = 1}^{2N-3} \log \det J_i
\end{equation}
with
\begin{equation}
    \log \det J_i = \log \frac{\partial \hat{z}_i}{\partial z_i} = \log \sigma_i^{-1} = - \log \sigma_i.
\end{equation}
In the reverse direction,
\begin{equation}
    z_i = \hat{z}_i \cdot \sigma_i + \mu_i
\end{equation}
and
\begin{equation}
    \log \det J_i^{-1} = \log \sigma_i.
\end{equation}

\subsection{Energy regularization}

Following the approach suggested in Ref.~\onlinecite{noe_boltzmann_2019_supp}, we regularize the potential energy of the system using the transformation
\begin{equation}
    U_{\text{polymer}}' =
    \begin{cases}
        U_{\text{polymer}} & \text{if} \, \, U_{\text{polymer}} < U_{\text{high}}, \\
        U_{\text{high}} + \log (U_{\text{polymer}} - U_{\text{high}} + 1) & \text{if} \, \, U_{\text{high}} \leq U_{\text{polymer}} < U_{\text{max}}, \\
        U_{\text{high}} + \log (U_{\text{max}} - U_{\text{high}} + 1) & \text{if} \, \, U_{\text{polymer}} \geq U_{\text{max}},
    \end{cases}
\end{equation}
where $U_{\text{high}}$ defines the onset of the energy regularization and $U_{\text{max}}$ the location of the function's plateau. If necessary, the value of $U_{\text{high}}$ can be varied as the training progresses.

\subsection{Training parameters}

\begin{table}[h]
    \centering
    \caption{Parameters used for testing GenAIMMD in the polymer model. In the case of the Boltzmann Generator, multiple values separated by commas indicate different parameters used inside a training scheduler.}
    \begin{tabular}{lccc}
        \toprule
        & \textbf{Boltzmann Generator} & \textbf{Committor model} & \textbf{GenAIMMD}\\
        \midrule
        Hidden layers & 3 & 3 & \\
        Nodes per layer & 200 & 64, 32, 16 & \\
        Epochs per cycle & 100, 50, 25 & 50 & \\
        Learning rate & $5 \cdot10^{-4}$, $10^{-4}$, $10^{-4}$ & $10^{-3}$ & \\
        Batch size & 250 & 150 & \\
        RealNVP blocks & 8 &  & \\
        $\lambda_{\text{KL}}$ & $0$, $10^{-4}$, $10^{-3}$ &  & \\
        $\lambda_{\text{ML}}$ & 1 &  & \\
        $U_{\text{high}}$ & 1 &  & \\
        $U_{\text{max}}$ & $10^{20}$ &  & \\
        \midrule
        Cycles & & & 40 \\
        Diffusion coefficient & & & 1.0 \\
        Timestep & & & $10^{-4}$ \\
        Relaxation MD steps & & & 3000 \\
        $k_{\text{bias}}$ & & & 2.5 \\
        FIFO queue size & & & $2 \cdot N_{\text{initial samples}}$ \\
        $N_{\text{initial samples}}$ & & & 6000 \\
        \bottomrule
    \end{tabular}
    \label{supp_tab:params_polymer}
\end{table}

The hyperparameters used for testing GenAIMMD in the polymer model and to define the Boltzmann Generator and the committor model are listed in Supplementary Table~\ref{supp_tab:params_polymer}. We use the same integration scheme as for the two-dimensional model, and its parameters are listed in Supplementary Table~\ref{supp_tab:params_polymer}.

\section{Benchmarking GenAIMMD against standard TPS}

\begin{figure}
    \centering
    \includegraphics[width=\textwidth]{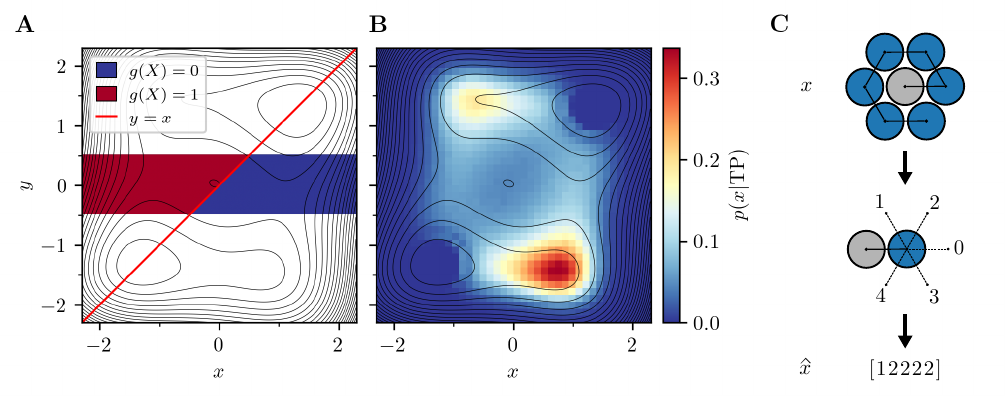}
    \caption{Details about the benchmarking of our method against conventional TPS. (A)~Regions contributing to the two possible values of the reaction channel indicator function $g(X)$ in the two-dimensional system, which is 1 for paths passing through the upper (red) channel and 0 for those using the lower channel (blue). (B) Reference $p(x | \mathrm{TP})$ histogram in the two-dimensional model, obtained from many de-correlated TPS trajectories. (C) Discretization of polymer configurations based on bond angles. Each bond angle is assigned a number (0 -- 4) based on the closest multiple of 60° (dashed lines), resulting in a 5-digit class for each configuration.}
    \label{supp_fig:benchmarking}
\end{figure}

In this section, we describe the steps taken to produce Fig.~4 in the main text, benchmarking conventional TPS against our method of obtaining transition paths. All dynamics were performed using the same integrator and parameters as described in each test system's section above.

\subsection{Reference and path sampling}

The presented benchmark data are based on different path sampling strategies. As reference for all metrics, we use $25\,000$ de-correlated transition paths from standard TPS (2D model) and a long-running TPS simulation split across $30$ workers, each one producing $30\,000$ paths with an output frequency of $15$, resulting in a total of $13.5$ million shooting trials and $900\,000$ transition paths (polymer model). For the performance analysis of standard TPS, we use 30 individual TPS runs with a path output frequency of 1 that produce $10\,000$ transition paths each for the two-dimensional model and the same number of workers and shooting trials as for the reference data in the case of the polymer model. All standard TPS simulations use a uniform shooting point selection probability. On the other hand, for the same analysis of the generated shooting point paths, we produce $30 \cdot 10\,000$ shooting points in both systems and divide them into 30 blocks to facilitate comparison to the 30 individual TPS runs.

\subsection{Path histograms and polymer discretization}

Configurations along transition paths are distributed according to the distribution $p(x | \mathrm{TP)}$, which is related to the committor via~\cite{hummer_transition_2004_supp, e_transition-path_2010_supp}
\begin{equation}
    p(x | \mathrm{TP}) \propto \rho(x) \cdot p_\mathrm{B}(x) \cdot (1 - p_\mathrm{B}(x)).
\end{equation}
In our benchmark, we test the convergence of the two methods towards this distribution by comparing the $L^1$ distances between the $p(x | \mathrm{TP)}$ histograms they produce after $n_\mathrm{trials}$ shooting attempts and the reference's converged histogram (Fig.~4, first row). For the two-dimensional model, this reference histogram is depicted in Supplementary Figure~\ref{supp_fig:benchmarking}B. It uses a grid of $40 \times 40$ square bins between $-2.3$ and $2.3$, resulting in a lattice constant of $a = 0.115$.

In the higher-dimensional polymer model, configuration space is discretized by assigning each polymer a class $\hat{x}$ based on its bond angles, following the approach of Ref.~\onlinecite{falkner_conditioning_2023_supp}. These classes then serve as bins for the histogram. A class is defined as a 5-digit number, where each digit can range from 0 to 4, resulting in a theoretical total of $5^5 = 3125$ polymer classes. Of those, many are impossible to populate because of particle overlap. Each digit in a class corresponds to one of the 5 bond angles of the polymer, and its value (0--4) is chosen based on the closest multiple of 60° (see Supplementary Figure~\ref{supp_fig:benchmarking}C). The histogram is symmetrized with respect to mirroring of polymer configurations, since the model is invariant under this transformation. This is achieved by mirroring each polymer if its first bond angle satisfies $\sin \varphi_1 < 0$.

\subsection{Reaction channel indicator function}

The two-dimensional model has two distinct reaction channels connecting the two stable states. To identify the channel followed by a path $X = \{(x_1, y_1), \dots, (x_{L(X)}, y_{L(X)})\}$, we use a path indicator function defined by
\begin{equation}
    g(X) = H \left[ \frac{\sum_{i = 1}^{L(X)} \mathds{1}_{|y_i| \, < \, 0.5} \, (y_i - x_i)}{\sum_{i = 1}^{L(X)} \mathds{1}_{|y_i| \, < \, 0.5}} \right] =
    \begin{cases}
        1 & \text{upper channel chosen,} \\
        0 & \text{lower channel chosen,}
    \end{cases}
\end{equation}
where $H(x)$ is the Heaviside step function and $\mathds{1}_\mathrm{condition}$ is the indicator function, which is 1 if the condition is met and 0 otherwise. In practice, $g(X)$ selects the channel in which a path $X$ spends most of its time when close to the barrier region. The two channels are defined by the colored regions in Supplementary Figure~\ref{supp_fig:benchmarking}A.


\begin{thebibliography}{54}%
\makeatletter
\providecommand \@ifxundefined [1]{%
 \@ifx{#1\undefined}
}%
\providecommand \@ifnum [1]{%
 \ifnum #1\expandafter \@firstoftwo
 \else \expandafter \@secondoftwo
 \fi
}%
\providecommand \@ifx [1]{%
 \ifx #1\expandafter \@firstoftwo
 \else \expandafter \@secondoftwo
 \fi
}%
\providecommand \natexlab [1]{#1}%
\providecommand \enquote  [1]{``#1''}%
\providecommand \bibnamefont  [1]{#1}%
\providecommand \bibfnamefont [1]{#1}%
\providecommand \citenamefont [1]{#1}%
\providecommand \href@noop [0]{\@secondoftwo}%
\providecommand \href [0]{\begingroup \@sanitize@url \@href}%
\providecommand \@href[1]{\@@startlink{#1}\@@href}%
\providecommand \@@href[1]{\endgroup#1\@@endlink}%
\providecommand \@sanitize@url [0]{\catcode `\\12\catcode `\$12\catcode `\&12\catcode `\#12\catcode `\^12\catcode `\_12\catcode `\%12\relax}%
\providecommand \@@startlink[1]{}%
\providecommand \@@endlink[0]{}%
\providecommand \url  [0]{\begingroup\@sanitize@url \@url }%
\providecommand \@url [1]{\endgroup\@href {#1}{\urlprefix }}%
\providecommand \urlprefix  [0]{URL }%
\providecommand \Eprint [0]{\href }%
\providecommand \doibase [0]{http://dx.doi.org/}%
\providecommand \selectlanguage [0]{\@gobble}%
\providecommand \bibinfo  [0]{\@secondoftwo}%
\providecommand \bibfield  [0]{\@secondoftwo}%
\providecommand \translation [1]{[#1]}%
\providecommand \BibitemOpen [0]{}%
\providecommand \bibitemStop [0]{}%
\providecommand \bibitemNoStop [0]{.\EOS\space}%
\providecommand \EOS [0]{\spacefactor3000\relax}%
\providecommand \BibitemShut  [1]{\csname bibitem#1\endcsname}%
\let\auto@bib@innerbib\@empty
\bibitem [{\citenamefont {Pan}\ and\ \citenamefont {Chandler}(2004)}]{pan_dynamics_2004}%
  \BibitemOpen
  \bibfield  {author} {\bibinfo {author} {\bibfnamefont {A.~C.}\ \bibnamefont {Pan}}\ and\ \bibinfo {author} {\bibfnamefont {D.}~\bibnamefont {Chandler}},\ }\bibfield  {title} {{\selectlanguage {en}\enquote {\bibinfo {title} {{Dynamics of Nucleation in the Ising Model}},}\ }}\href {\doibase 10.1021/jp0471249} {\bibfield  {journal} {\bibinfo  {journal} {The Journal of Physical Chemistry B}\ }\textbf {\bibinfo {volume} {108}},\ \bibinfo {pages} {19681--19686} (\bibinfo {year} {2004})}\BibitemShut {NoStop}%
\bibitem [{\citenamefont {Juraszek}\ and\ \citenamefont {Bolhuis}(2006)}]{juraszek_sampling_2006}%
  \BibitemOpen
  \bibfield  {author} {\bibinfo {author} {\bibfnamefont {J.}~\bibnamefont {Juraszek}}\ and\ \bibinfo {author} {\bibfnamefont {P.~G.}\ \bibnamefont {Bolhuis}},\ }\bibfield  {title} {{\selectlanguage {en}\enquote {\bibinfo {title} {Sampling the multiple folding mechanisms of {Trp}-cage in explicit solvent},}\ }}\href {\doibase 10.1073/pnas.0606692103} {\bibfield  {journal} {\bibinfo  {journal} {Proceedings of the National Academy of Sciences}\ }\textbf {\bibinfo {volume} {103}},\ \bibinfo {pages} {15859--15864} (\bibinfo {year} {2006})}\BibitemShut {NoStop}%
\bibitem [{\citenamefont {Okazaki}\ \emph {et~al.}(2019)\citenamefont {Okazaki}, \citenamefont {Wöhlert}, \citenamefont {Warnau}, \citenamefont {Jung}, \citenamefont {Yildiz}, \citenamefont {Kühlbrandt},\ and\ \citenamefont {Hummer}}]{okazaki_mechanism_2019}%
  \BibitemOpen
  \bibfield  {author} {\bibinfo {author} {\bibfnamefont {K.-i.}\ \bibnamefont {Okazaki}}, \bibinfo {author} {\bibfnamefont {D.}~\bibnamefont {Wöhlert}}, \bibinfo {author} {\bibfnamefont {J.}~\bibnamefont {Warnau}}, \bibinfo {author} {\bibfnamefont {H.}~\bibnamefont {Jung}}, \bibinfo {author} {\bibfnamefont {{\"O}.}~\bibnamefont {Yildiz}}, \bibinfo {author} {\bibfnamefont {W.}~\bibnamefont {Kühlbrandt}}, \ and\ \bibinfo {author} {\bibfnamefont {G.}~\bibnamefont {Hummer}},\ }\bibfield  {title} {{\selectlanguage {en}\enquote {\bibinfo {title} {Mechanism of the electroneutral sodium/proton antiporter {PaNhaP} from transition-path shooting},}\ }}\href {\doibase 10.1038/s41467-019-09739-0} {\bibfield  {journal} {\bibinfo  {journal} {Nature Communications}\ }\textbf {\bibinfo {volume} {10}},\ \bibinfo {pages} {1742} (\bibinfo {year} {2019})}\BibitemShut {NoStop}%
\bibitem [{\citenamefont {Angiolari}\ \emph {et~al.}(2025)\citenamefont {Angiolari}, \citenamefont {Coretti}, \citenamefont {Salanne},\ and\ \citenamefont {Bonella}}]{angiolari_electrically_2025}%
  \BibitemOpen
  \bibfield  {author} {\bibinfo {author} {\bibfnamefont {F.}~\bibnamefont {Angiolari}}, \bibinfo {author} {\bibfnamefont {A.}~\bibnamefont {Coretti}}, \bibinfo {author} {\bibfnamefont {M.}~\bibnamefont {Salanne}}, \ and\ \bibinfo {author} {\bibfnamefont {S.}~\bibnamefont {Bonella}},\ }\bibfield  {title} {{\selectlanguage {en}\enquote {\bibinfo {title} {Electrically driven first-order phase transition of a {2D} ionic crystal at the electrode/electrolyte interface},}\ }}\href {\doibase 10.1073/pnas.2520026122} {\bibfield  {journal} {\bibinfo  {journal} {Proceedings of the National Academy of Sciences}\ }\textbf {\bibinfo {volume} {122}},\ \bibinfo {pages} {e2520026122} (\bibinfo {year} {2025})}\BibitemShut {NoStop}%
\bibitem [{\citenamefont {Falkner}\ and\ \citenamefont {Schwierz}(2021)}]{falkner_kinetic_2021}%
  \BibitemOpen
  \bibfield  {author} {\bibinfo {author} {\bibfnamefont {S.}~\bibnamefont {Falkner}}\ and\ \bibinfo {author} {\bibfnamefont {N.}~\bibnamefont {Schwierz}},\ }\bibfield  {title} {{\selectlanguage {en}\enquote {\bibinfo {title} {Kinetic pathways of water exchange in the first hydration shell of magnesium: {Influence} of water model and ionic force field},}\ }}\href {\doibase 10.1063/5.0060896} {\bibfield  {journal} {\bibinfo  {journal} {The Journal of Chemical Physics}\ }\textbf {\bibinfo {volume} {155}},\ \bibinfo {pages} {084503} (\bibinfo {year} {2021})}\BibitemShut {NoStop}%
\bibitem [{\citenamefont {Hénin}\ \emph {et~al.}(2022)\citenamefont {Hénin}, \citenamefont {Lelièvre}, \citenamefont {Shirts}, \citenamefont {Valsson},\ and\ \citenamefont {Delemotte}}]{henin_enhanced_2022}%
  \BibitemOpen
  \bibfield  {author} {\bibinfo {author} {\bibfnamefont {J.}~\bibnamefont {Hénin}}, \bibinfo {author} {\bibfnamefont {T.}~\bibnamefont {Lelièvre}}, \bibinfo {author} {\bibfnamefont {M.~R.}\ \bibnamefont {Shirts}}, \bibinfo {author} {\bibfnamefont {O.}~\bibnamefont {Valsson}}, \ and\ \bibinfo {author} {\bibfnamefont {L.}~\bibnamefont {Delemotte}},\ }\bibfield  {title} {\enquote {\bibinfo {title} {Enhanced sampling methods for molecular dynamics simulations},}\ }\href {\doibase 10.33011/livecoms.4.1.1583} {\bibfield  {journal} {\bibinfo  {journal} {Living Journal of Computational Molecular Science}\ }\textbf {\bibinfo {volume} {4}} (\bibinfo {year} {2022}),\ 10.33011/livecoms.4.1.1583},\ \bibinfo {note} {arXiv:2202.04164 [cond-mat]}\BibitemShut {NoStop}%
\bibitem [{\citenamefont {Torrie}\ and\ \citenamefont {Valleau}(1977)}]{torrie_nonphysical_1977}%
  \BibitemOpen
  \bibfield  {author} {\bibinfo {author} {\bibfnamefont {G.}~\bibnamefont {Torrie}}\ and\ \bibinfo {author} {\bibfnamefont {J.}~\bibnamefont {Valleau}},\ }\bibfield  {title} {{\selectlanguage {en}\enquote {\bibinfo {title} {Nonphysical sampling distributions in {Monte} {Carlo} free-energy estimation: {Umbrella} sampling},}\ }}\href {\doibase 10.1016/0021-9991(77)90121-8} {\bibfield  {journal} {\bibinfo  {journal} {Journal of Computational Physics}\ }\textbf {\bibinfo {volume} {23}},\ \bibinfo {pages} {187--199} (\bibinfo {year} {1977})}\BibitemShut {NoStop}%
\bibitem [{\citenamefont {Kästner}(2011)}]{kastner_umbrella_2011}%
  \BibitemOpen
  \bibfield  {author} {\bibinfo {author} {\bibfnamefont {J.}~\bibnamefont {Kästner}},\ }\bibfield  {title} {{\selectlanguage {en}\enquote {\bibinfo {title} {Umbrella sampling},}\ }}\href {\doibase 10.1002/wcms.66} {\bibfield  {journal} {\bibinfo  {journal} {WIREs Computational Molecular Science}\ }\textbf {\bibinfo {volume} {1}},\ \bibinfo {pages} {932--942} (\bibinfo {year} {2011})}\BibitemShut {NoStop}%
\bibitem [{\citenamefont {Laio}\ and\ \citenamefont {Parrinello}(2002)}]{laio_escaping_2002}%
  \BibitemOpen
  \bibfield  {author} {\bibinfo {author} {\bibfnamefont {A.}~\bibnamefont {Laio}}\ and\ \bibinfo {author} {\bibfnamefont {M.}~\bibnamefont {Parrinello}},\ }\bibfield  {title} {{\selectlanguage {en}\enquote {\bibinfo {title} {Escaping free-energy minima},}\ }}\href {\doibase 10.1073/pnas.202427399} {\bibfield  {journal} {\bibinfo  {journal} {Proceedings of the National Academy of Sciences}\ }\textbf {\bibinfo {volume} {99}},\ \bibinfo {pages} {12562--12566} (\bibinfo {year} {2002})}\BibitemShut {NoStop}%
\bibitem [{\citenamefont {Dellago}\ and\ \citenamefont {Bolhuis}(2009)}]{dellago_transition_2009}%
  \BibitemOpen
  \bibfield  {author} {\bibinfo {author} {\bibfnamefont {C.}~\bibnamefont {Dellago}}\ and\ \bibinfo {author} {\bibfnamefont {P.~G.}\ \bibnamefont {Bolhuis}},\ }\enquote {\bibinfo {title} {Transition path sampling and other advanced simulation techniques for rare events},}\ in\ \href {\doibase 10.1007/978-3-540-87706-6_3} {\emph {\bibinfo {booktitle} {Advanced Computer Simulation Approaches for Soft Matter Sciences III}}},\ \bibinfo {editor} {edited by\ \bibinfo {editor} {\bibfnamefont {C.}~\bibnamefont {Holm}}\ and\ \bibinfo {editor} {\bibfnamefont {K.}~\bibnamefont {Kremer}}}\ (\bibinfo  {publisher} {Springer Berlin Heidelberg},\ \bibinfo {address} {Berlin, Heidelberg},\ \bibinfo {year} {2009})\ pp.\ \bibinfo {pages} {167--233}\BibitemShut {NoStop}%
\bibitem [{\citenamefont {Ma}\ and\ \citenamefont {Dinner}(2005)}]{ma_automatic_2005}%
  \BibitemOpen
  \bibfield  {author} {\bibinfo {author} {\bibfnamefont {A.}~\bibnamefont {Ma}}\ and\ \bibinfo {author} {\bibfnamefont {A.~R.}\ \bibnamefont {Dinner}},\ }\bibfield  {title} {{\selectlanguage {en}\enquote {\bibinfo {title} {Automatic {Method} for {Identifying} {Reaction} {Coordinates} in {Complex} {Systems}},}\ }}\href {\doibase 10.1021/jp045546c} {\bibfield  {journal} {\bibinfo  {journal} {The Journal of Physical Chemistry B}\ }\textbf {\bibinfo {volume} {109}},\ \bibinfo {pages} {6769--6779} (\bibinfo {year} {2005})}\BibitemShut {NoStop}%
\bibitem [{\citenamefont {Peters}\ and\ \citenamefont {Trout}(2006)}]{peters_obtaining_2006}%
  \BibitemOpen
  \bibfield  {author} {\bibinfo {author} {\bibfnamefont {B.}~\bibnamefont {Peters}}\ and\ \bibinfo {author} {\bibfnamefont {B.~L.}\ \bibnamefont {Trout}},\ }\bibfield  {title} {{\selectlanguage {en}\enquote {\bibinfo {title} {Obtaining reaction coordinates by likelihood maximization},}\ }}\href {\doibase 10.1063/1.2234477} {\bibfield  {journal} {\bibinfo  {journal} {The Journal of Chemical Physics}\ }\textbf {\bibinfo {volume} {125}},\ \bibinfo {pages} {054108} (\bibinfo {year} {2006})}\BibitemShut {NoStop}%
\bibitem [{\citenamefont {Jung}\ \emph {et~al.}(2023)\citenamefont {Jung}, \citenamefont {Covino}, \citenamefont {Arjun}, \citenamefont {Leitold}, \citenamefont {Dellago}, \citenamefont {Bolhuis},\ and\ \citenamefont {Hummer}}]{jung_machine-guided_2023}%
  \BibitemOpen
  \bibfield  {author} {\bibinfo {author} {\bibfnamefont {H.}~\bibnamefont {Jung}}, \bibinfo {author} {\bibfnamefont {R.}~\bibnamefont {Covino}}, \bibinfo {author} {\bibfnamefont {A.}~\bibnamefont {Arjun}}, \bibinfo {author} {\bibfnamefont {C.}~\bibnamefont {Leitold}}, \bibinfo {author} {\bibfnamefont {C.}~\bibnamefont {Dellago}}, \bibinfo {author} {\bibfnamefont {P.~G.}\ \bibnamefont {Bolhuis}}, \ and\ \bibinfo {author} {\bibfnamefont {G.}~\bibnamefont {Hummer}},\ }\bibfield  {title} {{\selectlanguage {en}\enquote {\bibinfo {title} {Machine-guided path sampling to discover mechanisms of molecular self-organization},}\ }}\href {\doibase 10.1038/s43588-023-00428-z} {\bibfield  {journal} {\bibinfo  {journal} {Nature Computational Science}\ }\textbf {\bibinfo {volume} {3}},\ \bibinfo {pages} {334--345} (\bibinfo {year} {2023})}\BibitemShut {NoStop}%
\bibitem [{\citenamefont {Kang}, \citenamefont {Trizio},\ and\ \citenamefont {Parrinello}(2024)}]{kang_computing_2024}%
  \BibitemOpen
  \bibfield  {author} {\bibinfo {author} {\bibfnamefont {P.}~\bibnamefont {Kang}}, \bibinfo {author} {\bibfnamefont {E.}~\bibnamefont {Trizio}}, \ and\ \bibinfo {author} {\bibfnamefont {M.}~\bibnamefont {Parrinello}},\ }\bibfield  {title} {{\selectlanguage {en}\enquote {\bibinfo {title} {Computing the committor with the committor to study the transition state ensemble},}\ }}\href {\doibase 10.1038/s43588-024-00645-0} {\bibfield  {journal} {\bibinfo  {journal} {Nature Computational Science}\ }\textbf {\bibinfo {volume} {4}},\ \bibinfo {pages} {451--460} (\bibinfo {year} {2024})}\BibitemShut {NoStop}%
\bibitem [{\citenamefont {Megías}\ \emph {et~al.}(2025)\citenamefont {Megías}, \citenamefont {Contreras~Arredondo}, \citenamefont {Chen}, \citenamefont {Tang}, \citenamefont {Roux},\ and\ \citenamefont {Chipot}}]{megias_iterative_2025}%
  \BibitemOpen
  \bibfield  {author} {\bibinfo {author} {\bibfnamefont {A.}~\bibnamefont {Megías}}, \bibinfo {author} {\bibfnamefont {S.}~\bibnamefont {Contreras~Arredondo}}, \bibinfo {author} {\bibfnamefont {C.~G.}\ \bibnamefont {Chen}}, \bibinfo {author} {\bibfnamefont {C.}~\bibnamefont {Tang}}, \bibinfo {author} {\bibfnamefont {B.}~\bibnamefont {Roux}}, \ and\ \bibinfo {author} {\bibfnamefont {C.}~\bibnamefont {Chipot}},\ }\bibfield  {title} {{\selectlanguage {en}\enquote {\bibinfo {title} {Iterative variational learning of committor-consistent transition pathways using artificial neural networks},}\ }}\href {\doibase 10.1038/s43588-025-00828-3} {\bibfield  {journal} {\bibinfo  {journal} {Nature Computational Science}\ }\textbf {\bibinfo {volume} {5}},\ \bibinfo {pages} {592--602} (\bibinfo {year} {2025})}\BibitemShut {NoStop}%
\bibitem [{\citenamefont {Falkner}\ \emph {et~al.}(2023)\citenamefont {Falkner}, \citenamefont {Coretti}, \citenamefont {Romano}, \citenamefont {Geissler},\ and\ \citenamefont {Dellago}}]{falkner_conditioning_2023}%
  \BibitemOpen
  \bibfield  {author} {\bibinfo {author} {\bibfnamefont {S.}~\bibnamefont {Falkner}}, \bibinfo {author} {\bibfnamefont {A.}~\bibnamefont {Coretti}}, \bibinfo {author} {\bibfnamefont {S.}~\bibnamefont {Romano}}, \bibinfo {author} {\bibfnamefont {P.~L.}\ \bibnamefont {Geissler}}, \ and\ \bibinfo {author} {\bibfnamefont {C.}~\bibnamefont {Dellago}},\ }\bibfield  {title} {\enquote {\bibinfo {title} {Conditioning {Boltzmann} generators for rare event sampling},}\ }\href {\doibase 10.1088/2632-2153/acf55c} {\bibfield  {journal} {\bibinfo  {journal} {Machine Learning: Science and Technology}\ }\textbf {\bibinfo {volume} {4}},\ \bibinfo {pages} {035050} (\bibinfo {year} {2023})}\BibitemShut {NoStop}%
\bibitem [{\citenamefont {Asghar}\ \emph {et~al.}(2024)\citenamefont {Asghar}, \citenamefont {Pei}, \citenamefont {Volpe},\ and\ \citenamefont {Ni}}]{asghar_efficient_2024}%
  \BibitemOpen
  \bibfield  {author} {\bibinfo {author} {\bibfnamefont {S.}~\bibnamefont {Asghar}}, \bibinfo {author} {\bibfnamefont {Q.-X.}\ \bibnamefont {Pei}}, \bibinfo {author} {\bibfnamefont {G.}~\bibnamefont {Volpe}}, \ and\ \bibinfo {author} {\bibfnamefont {R.}~\bibnamefont {Ni}},\ }\bibfield  {title} {{\selectlanguage {en}\enquote {\bibinfo {title} {Efficient rare event sampling with unsupervised normalizing flows},}\ }}\href {\doibase 10.1038/s42256-024-00918-3} {\bibfield  {journal} {\bibinfo  {journal} {Nature Machine Intelligence}\ }\textbf {\bibinfo {volume} {6}},\ \bibinfo {pages} {1370--1381} (\bibinfo {year} {2024})}\BibitemShut {NoStop}%
\bibitem [{\citenamefont {Li}\ \emph {et~al.}(2026)\citenamefont {Li}, \citenamefont {Chen}, \citenamefont {Zhang},\ and\ \citenamefont {Pan}}]{li_differentiable_2026}%
  \BibitemOpen
  \bibfield  {author} {\bibinfo {author} {\bibfnamefont {S.-H.}\ \bibnamefont {Li}}, \bibinfo {author} {\bibfnamefont {C.}~\bibnamefont {Chen}}, \bibinfo {author} {\bibfnamefont {Y.-W.}\ \bibnamefont {Zhang}}, \ and\ \bibinfo {author} {\bibfnamefont {D.}~\bibnamefont {Pan}},\ }\href {\doibase 10.48550/arXiv.2604.09769} {{\selectlanguage {en}\enquote {\bibinfo {title} {Differentiable free energy surface: a variational approach to directly observing rare events using generative deep-learning models},}\ }} (\bibinfo {year} {2026}),\ \bibinfo {note} {arXiv:2604.09769 [physics.comp-ph]}\BibitemShut {NoStop}%
\bibitem [{\citenamefont {Tang}\ \emph {et~al.}(2026)\citenamefont {Tang}, \citenamefont {Pandey}, \citenamefont {Chen}, \citenamefont {Megías}, \citenamefont {Dehez},\ and\ \citenamefont {Chipot}}]{tang_breaking_2026}%
  \BibitemOpen
  \bibfield  {author} {\bibinfo {author} {\bibfnamefont {C.}~\bibnamefont {Tang}}, \bibinfo {author} {\bibfnamefont {M.~P.}\ \bibnamefont {Pandey}}, \bibinfo {author} {\bibfnamefont {C.~G.}\ \bibnamefont {Chen}}, \bibinfo {author} {\bibfnamefont {A.}~\bibnamefont {Megías}}, \bibinfo {author} {\bibfnamefont {F.}~\bibnamefont {Dehez}}, \ and\ \bibinfo {author} {\bibfnamefont {C.}~\bibnamefont {Chipot}},\ }\bibfield  {title} {{\selectlanguage {en}\enquote {\bibinfo {title} {Breaking timescales with generative sampling of conformational transitions},}\ }}\href {\doibase 10.1038/s41586-026-11025-1} {\bibfield  {journal} {\bibinfo  {journal} {Nature}\ } (\bibinfo {year} {2026}),\ 10.1038/s41586-026-11025-1}\BibitemShut {NoStop}%
\bibitem [{\citenamefont {Dellago}\ \emph {et~al.}(1998)\citenamefont {Dellago}, \citenamefont {Bolhuis}, \citenamefont {Csajka},\ and\ \citenamefont {Chandler}}]{dellago_transition_1998}%
  \BibitemOpen
  \bibfield  {author} {\bibinfo {author} {\bibfnamefont {C.}~\bibnamefont {Dellago}}, \bibinfo {author} {\bibfnamefont {P.~G.}\ \bibnamefont {Bolhuis}}, \bibinfo {author} {\bibfnamefont {F.~S.}\ \bibnamefont {Csajka}}, \ and\ \bibinfo {author} {\bibfnamefont {D.}~\bibnamefont {Chandler}},\ }\bibfield  {title} {{\selectlanguage {en}\enquote {\bibinfo {title} {Transition path sampling and the calculation of rate constants},}\ }}\href {\doibase 10.1063/1.475562} {\bibfield  {journal} {\bibinfo  {journal} {The Journal of Chemical Physics}\ }\textbf {\bibinfo {volume} {108}},\ \bibinfo {pages} {1964--1977} (\bibinfo {year} {1998})}\BibitemShut {NoStop}%
\bibitem [{\citenamefont {Bolhuis}\ \emph {et~al.}(2002)\citenamefont {Bolhuis}, \citenamefont {Chandler}, \citenamefont {Dellago},\ and\ \citenamefont {Geissler}}]{bolhuis_transition_2002}%
  \BibitemOpen
  \bibfield  {author} {\bibinfo {author} {\bibfnamefont {P.~G.}\ \bibnamefont {Bolhuis}}, \bibinfo {author} {\bibfnamefont {D.}~\bibnamefont {Chandler}}, \bibinfo {author} {\bibfnamefont {C.}~\bibnamefont {Dellago}}, \ and\ \bibinfo {author} {\bibfnamefont {P.~L.}\ \bibnamefont {Geissler}},\ }\bibfield  {title} {{\selectlanguage {en}\enquote {\bibinfo {title} {Transition {Path} {Sampling}: {Throwing} {Ropes} {Over} {Rough} {Mountain} {Passes}, in the {Dark}},}\ }}\href {\doibase 10.1146/annurev.physchem.53.082301.113146} {\bibfield  {journal} {\bibinfo  {journal} {Annual Review of Physical Chemistry}\ }\textbf {\bibinfo {volume} {53}},\ \bibinfo {pages} {291--318} (\bibinfo {year} {2002})}\BibitemShut {NoStop}%
\bibitem [{\citenamefont {Dellago}, \citenamefont {Bolhuis},\ and\ \citenamefont {Chandler}(1998)}]{dellago_efficient_1998}%
  \BibitemOpen
  \bibfield  {author} {\bibinfo {author} {\bibfnamefont {C.}~\bibnamefont {Dellago}}, \bibinfo {author} {\bibfnamefont {P.~G.}\ \bibnamefont {Bolhuis}}, \ and\ \bibinfo {author} {\bibfnamefont {D.}~\bibnamefont {Chandler}},\ }\bibfield  {title} {{\selectlanguage {en}\enquote {\bibinfo {title} {Efficient transition path sampling: {Application} to {Lennard}-{Jones} cluster rearrangements},}\ }}\href {\doibase 10.1063/1.476378} {\bibfield  {journal} {\bibinfo  {journal} {The Journal of Chemical Physics}\ }\textbf {\bibinfo {volume} {108}},\ \bibinfo {pages} {9236--9245} (\bibinfo {year} {1998})}\BibitemShut {NoStop}%
\bibitem [{\citenamefont {Jung}, \citenamefont {Okazaki},\ and\ \citenamefont {Hummer}(2017)}]{jung_transition_2017}%
  \BibitemOpen
  \bibfield  {author} {\bibinfo {author} {\bibfnamefont {H.}~\bibnamefont {Jung}}, \bibinfo {author} {\bibfnamefont {K.-i.}\ \bibnamefont {Okazaki}}, \ and\ \bibinfo {author} {\bibfnamefont {G.}~\bibnamefont {Hummer}},\ }\bibfield  {title} {{\selectlanguage {en}\enquote {\bibinfo {title} {Transition path sampling of rare events by shooting from the top},}\ }}\href {\doibase 10.1063/1.4997378} {\bibfield  {journal} {\bibinfo  {journal} {The Journal of Chemical Physics}\ }\textbf {\bibinfo {volume} {147}},\ \bibinfo {pages} {152716} (\bibinfo {year} {2017})}\BibitemShut {NoStop}%
\bibitem [{\citenamefont {Bolhuis}(2003{\natexlab{a}})}]{bolhuis_transition_2003}%
  \BibitemOpen
  \bibfield  {author} {\bibinfo {author} {\bibfnamefont {P.~G.}\ \bibnamefont {Bolhuis}},\ }\bibfield  {title} {{\selectlanguage {en}\enquote {\bibinfo {title} {Transition path sampling on diffusive barriers},}\ }}\href {\doibase 10.1088/0953-8984/15/1/314} {\bibfield  {journal} {\bibinfo  {journal} {Journal of Physics: Condensed Matter}\ }\textbf {\bibinfo {volume} {15}},\ \bibinfo {pages} {S113--S120} (\bibinfo {year} {2003}{\natexlab{a}})}\BibitemShut {NoStop}%
\bibitem [{\citenamefont {Bolhuis}(2003{\natexlab{b}})}]{bolhuis_transition-path_2003}%
  \BibitemOpen
  \bibfield  {author} {\bibinfo {author} {\bibfnamefont {P.~G.}\ \bibnamefont {Bolhuis}},\ }\bibfield  {title} {{\selectlanguage {en}\enquote {\bibinfo {title} {Transition-path sampling of $\beta$-hairpin folding},}\ }}\href {\doibase 10.1073/pnas.1534924100} {\bibfield  {journal} {\bibinfo  {journal} {Proceedings of the National Academy of Sciences}\ }\textbf {\bibinfo {volume} {100}},\ \bibinfo {pages} {12129--12134} (\bibinfo {year} {2003}{\natexlab{b}})}\BibitemShut {NoStop}%
\bibitem [{\citenamefont {Menzl}, \citenamefont {Singraber},\ and\ \citenamefont {Dellago}(2016)}]{menzl_s-shooting_2016}%
  \BibitemOpen
  \bibfield  {author} {\bibinfo {author} {\bibfnamefont {G.}~\bibnamefont {Menzl}}, \bibinfo {author} {\bibfnamefont {A.}~\bibnamefont {Singraber}}, \ and\ \bibinfo {author} {\bibfnamefont {C.}~\bibnamefont {Dellago}},\ }\bibfield  {title} {{\selectlanguage {en}\enquote {\bibinfo {title} {S-shooting: a {Bennett}–{Chandler}-like method for the computation of rate constants from committor trajectories},}\ }}\href {\doibase 10.1039/C6FD00124F} {\bibfield  {journal} {\bibinfo  {journal} {Faraday Discussions}\ }\textbf {\bibinfo {volume} {195}},\ \bibinfo {pages} {345--364} (\bibinfo {year} {2016})}\BibitemShut {NoStop}%
\bibitem [{\citenamefont {Bolhuis}\ and\ \citenamefont {Swenson}(2021)}]{bolhuis_transition_2021}%
  \BibitemOpen
  \bibfield  {author} {\bibinfo {author} {\bibfnamefont {P.~G.}\ \bibnamefont {Bolhuis}}\ and\ \bibinfo {author} {\bibfnamefont {D.~W.~H.}\ \bibnamefont {Swenson}},\ }\bibfield  {title} {{\selectlanguage {en}\enquote {\bibinfo {title} {Transition {Path} {Sampling} as {Markov} {Chain} {Monte} {Carlo} of {Trajectories}: {Recent} {Algorithms}, {Software}, {Applications}, and {Future} {Outlook}},}\ }}\href {\doibase 10.1002/adts.202000237} {\bibfield  {journal} {\bibinfo  {journal} {Advanced Theory and Simulations}\ }\textbf {\bibinfo {volume} {4}},\ \bibinfo {pages} {2000237} (\bibinfo {year} {2021})}\BibitemShut {NoStop}%
\bibitem [{\citenamefont {Onsager}(1938)}]{onsager_initial_1938}%
  \BibitemOpen
  \bibfield  {author} {\bibinfo {author} {\bibfnamefont {L.}~\bibnamefont {Onsager}},\ }\bibfield  {title} {{\selectlanguage {en}\enquote {\bibinfo {title} {Initial {Recombination} of {Ions}},}\ }}\href {\doibase 10.1103/PhysRev.54.554} {\bibfield  {journal} {\bibinfo  {journal} {Physical Review}\ }\textbf {\bibinfo {volume} {54}},\ \bibinfo {pages} {554--557} (\bibinfo {year} {1938})}\BibitemShut {NoStop}%
\bibitem [{\citenamefont {Du}\ \emph {et~al.}(1998)\citenamefont {Du}, \citenamefont {Pande}, \citenamefont {Grosberg}, \citenamefont {Tanaka},\ and\ \citenamefont {Shakhnovich}}]{du_transition_1998}%
  \BibitemOpen
  \bibfield  {author} {\bibinfo {author} {\bibfnamefont {R.}~\bibnamefont {Du}}, \bibinfo {author} {\bibfnamefont {V.~S.}\ \bibnamefont {Pande}}, \bibinfo {author} {\bibfnamefont {A.~Y.}\ \bibnamefont {Grosberg}}, \bibinfo {author} {\bibfnamefont {T.}~\bibnamefont {Tanaka}}, \ and\ \bibinfo {author} {\bibfnamefont {E.~S.}\ \bibnamefont {Shakhnovich}},\ }\bibfield  {title} {{\selectlanguage {en}\enquote {\bibinfo {title} {On the transition coordinate for protein folding},}\ }}\href {\doibase 10.1063/1.475393} {\bibfield  {journal} {\bibinfo  {journal} {The Journal of Chemical Physics}\ }\textbf {\bibinfo {volume} {108}},\ \bibinfo {pages} {334--350} (\bibinfo {year} {1998})}\BibitemShut {NoStop}%
\bibitem [{\citenamefont {Hummer}(2004)}]{hummer_transition_2004}%
  \BibitemOpen
  \bibfield  {author} {\bibinfo {author} {\bibfnamefont {G.}~\bibnamefont {Hummer}},\ }\bibfield  {title} {{\selectlanguage {en}\enquote {\bibinfo {title} {From transition paths to transition states and rate coefficients},}\ }}\href {\doibase 10.1063/1.1630572} {\bibfield  {journal} {\bibinfo  {journal} {The Journal of Chemical Physics}\ }\textbf {\bibinfo {volume} {120}},\ \bibinfo {pages} {516--523} (\bibinfo {year} {2004})}\BibitemShut {NoStop}%
\bibitem [{\citenamefont {E}\ and\ \citenamefont {Vanden-Eijnden}(2010)}]{e_transition-path_2010}%
  \BibitemOpen
  \bibfield  {author} {\bibinfo {author} {\bibfnamefont {W.}~\bibnamefont {E}}\ and\ \bibinfo {author} {\bibfnamefont {E.}~\bibnamefont {Vanden-Eijnden}},\ }\bibfield  {title} {{\selectlanguage {en}\enquote {\bibinfo {title} {Transition-{Path} {Theory} and {Path}-{Finding} {Algorithms} for the {Study} of {Rare} {Events}},}\ }}\href {\doibase 10.1146/annurev.physchem.040808.090412} {\bibfield  {journal} {\bibinfo  {journal} {Annual Review of Physical Chemistry}\ }\textbf {\bibinfo {volume} {61}},\ \bibinfo {pages} {391--420} (\bibinfo {year} {2010})}\BibitemShut {NoStop}%
\bibitem [{\citenamefont {Noé}\ \emph {et~al.}(2019)\citenamefont {Noé}, \citenamefont {Olsson}, \citenamefont {Köhler},\ and\ \citenamefont {Wu}}]{noe_boltzmann_2019}%
  \BibitemOpen
  \bibfield  {author} {\bibinfo {author} {\bibfnamefont {F.}~\bibnamefont {Noé}}, \bibinfo {author} {\bibfnamefont {S.}~\bibnamefont {Olsson}}, \bibinfo {author} {\bibfnamefont {J.}~\bibnamefont {Köhler}}, \ and\ \bibinfo {author} {\bibfnamefont {H.}~\bibnamefont {Wu}},\ }\bibfield  {title} {{\selectlanguage {en}\enquote {\bibinfo {title} {Boltzmann generators: {Sampling} equilibrium states of many-body systems with deep learning},}\ }}\href {\doibase 10.1126/science.aaw1147} {\bibfield  {journal} {\bibinfo  {journal} {Science}\ }\textbf {\bibinfo {volume} {365}},\ \bibinfo {pages} {eaaw1147} (\bibinfo {year} {2019})}\BibitemShut {NoStop}%
\bibitem [{\citenamefont {Tabak}\ and\ \citenamefont {Vanden-Eijnden}(2010)}]{tabak_density_2010}%
  \BibitemOpen
  \bibfield  {author} {\bibinfo {author} {\bibfnamefont {E.~G.}\ \bibnamefont {Tabak}}\ and\ \bibinfo {author} {\bibfnamefont {E.}~\bibnamefont {Vanden-Eijnden}},\ }\bibfield  {title} {\enquote {\bibinfo {title} {Density estimation by dual ascent of the log-likelihood},}\ }\href {https://projecteuclid.org/journals/communications-in-mathematical-sciences/volume-8/issue-1/Density-estimation-by-dual-ascent-of-the-log-likelihood/cms/1266935020.full} {\bibfield  {journal} {\bibinfo  {journal} {Communications in Mathematical Sciences}\ }\textbf {\bibinfo {volume} {8}},\ \bibinfo {pages} {217--233} (\bibinfo {year} {2010})}\BibitemShut {NoStop}%
\bibitem [{\citenamefont {Tabak}\ and\ \citenamefont {Turner}(2013)}]{tabak_family_2013}%
  \BibitemOpen
  \bibfield  {author} {\bibinfo {author} {\bibfnamefont {E.~G.}\ \bibnamefont {Tabak}}\ and\ \bibinfo {author} {\bibfnamefont {C.~V.}\ \bibnamefont {Turner}},\ }\bibfield  {title} {{\selectlanguage {en}\enquote {\bibinfo {title} {A {Family} of {Nonparametric} {Density} {Estimation} {Algorithms}},}\ }}\href {\doibase 10.1002/cpa.21423} {\bibfield  {journal} {\bibinfo  {journal} {Communications on Pure and Applied Mathematics}\ }\textbf {\bibinfo {volume} {66}},\ \bibinfo {pages} {145--164} (\bibinfo {year} {2013})}\BibitemShut {NoStop}%
\bibitem [{\citenamefont {Dinh}, \citenamefont {Sohl-Dickstein},\ and\ \citenamefont {Bengio}(2017)}]{dinh_density_2017}%
  \BibitemOpen
  \bibfield  {author} {\bibinfo {author} {\bibfnamefont {L.}~\bibnamefont {Dinh}}, \bibinfo {author} {\bibfnamefont {J.}~\bibnamefont {Sohl-Dickstein}}, \ and\ \bibinfo {author} {\bibfnamefont {S.}~\bibnamefont {Bengio}},\ }\href {\doibase 10.48550/arXiv.1605.08803} {\enquote {\bibinfo {title} {Density estimation using {Real} {NVP}},}\ } (\bibinfo {year} {2017}),\ \bibinfo {note} {arXiv:1605.08803}\BibitemShut {NoStop}%
\bibitem [{\citenamefont {Coretti}\ \emph {et~al.}(2024)\citenamefont {Coretti}, \citenamefont {Falkner}, \citenamefont {Weinreich}, \citenamefont {Dellago},\ and\ \citenamefont {von Lilienfeld}}]{coretti_boltzmann_2024}%
  \BibitemOpen
  \bibfield  {author} {\bibinfo {author} {\bibfnamefont {A.}~\bibnamefont {Coretti}}, \bibinfo {author} {\bibfnamefont {S.}~\bibnamefont {Falkner}}, \bibinfo {author} {\bibfnamefont {J.}~\bibnamefont {Weinreich}}, \bibinfo {author} {\bibfnamefont {C.}~\bibnamefont {Dellago}}, \ and\ \bibinfo {author} {\bibfnamefont {O.~A.}\ \bibnamefont {von Lilienfeld}},\ }\bibfield  {title} {\enquote {\bibinfo {title} {Boltzmann {Generators} and the {New} {Frontier} of {Computational} {Sampling} in {Many}-{Body} {Systems}},}\ }\href {\doibase 10.25950/bfa99422} {\bibfield  {journal} {\bibinfo  {journal} {KIM REVIEW}\ }\textbf {\bibinfo {volume} {2}} (\bibinfo {year} {2024}),\ 10.25950/bfa99422}\BibitemShut {NoStop}%
\bibitem [{\citenamefont {Wirnsberger}\ \emph {et~al.}(2020)\citenamefont {Wirnsberger}, \citenamefont {Ballard}, \citenamefont {Papamakarios}, \citenamefont {Abercrombie}, \citenamefont {Racanière}, \citenamefont {Pritzel}, \citenamefont {Jimenez~Rezende},\ and\ \citenamefont {Blundell}}]{wirnsberger_targeted_2020}%
  \BibitemOpen
  \bibfield  {author} {\bibinfo {author} {\bibfnamefont {P.}~\bibnamefont {Wirnsberger}}, \bibinfo {author} {\bibfnamefont {A.~J.}\ \bibnamefont {Ballard}}, \bibinfo {author} {\bibfnamefont {G.}~\bibnamefont {Papamakarios}}, \bibinfo {author} {\bibfnamefont {S.}~\bibnamefont {Abercrombie}}, \bibinfo {author} {\bibfnamefont {S.}~\bibnamefont {Racanière}}, \bibinfo {author} {\bibfnamefont {A.}~\bibnamefont {Pritzel}}, \bibinfo {author} {\bibfnamefont {D.}~\bibnamefont {Jimenez~Rezende}}, \ and\ \bibinfo {author} {\bibfnamefont {C.}~\bibnamefont {Blundell}},\ }\bibfield  {title} {{\selectlanguage {en}\enquote {\bibinfo {title} {Targeted free energy estimation via learned mappings},}\ }}\href {\doibase 10.1063/5.0018903} {\bibfield  {journal} {\bibinfo  {journal} {The Journal of Chemical Physics}\ }\textbf {\bibinfo {volume} {153}},\ \bibinfo {pages} {144112} (\bibinfo {year} {2020})}\BibitemShut {NoStop}%
\bibitem [{\citenamefont {Ahmad}\ and\ \citenamefont {Cai}(2022)}]{ahmad_free_2022}%
  \BibitemOpen
  \bibfield  {author} {\bibinfo {author} {\bibfnamefont {R.}~\bibnamefont {Ahmad}}\ and\ \bibinfo {author} {\bibfnamefont {W.}~\bibnamefont {Cai}},\ }\bibfield  {title} {{\selectlanguage {en}\enquote {\bibinfo {title} {Free energy calculation of crystalline solids using normalizing flows},}\ }}\href {\doibase 10.1088/1361-651X/ac7f4b} {\bibfield  {journal} {\bibinfo  {journal} {Modelling and Simulation in Materials Science and Engineering}\ }\textbf {\bibinfo {volume} {30}},\ \bibinfo {pages} {065007} (\bibinfo {year} {2022})}\BibitemShut {NoStop}%
\bibitem [{\citenamefont {Schebek}\ \emph {et~al.}(2024)\citenamefont {Schebek}, \citenamefont {Invernizzi}, \citenamefont {Noé},\ and\ \citenamefont {Rogal}}]{schebek_efficient_2024}%
  \BibitemOpen
  \bibfield  {author} {\bibinfo {author} {\bibfnamefont {M.}~\bibnamefont {Schebek}}, \bibinfo {author} {\bibfnamefont {M.}~\bibnamefont {Invernizzi}}, \bibinfo {author} {\bibfnamefont {F.}~\bibnamefont {Noé}}, \ and\ \bibinfo {author} {\bibfnamefont {J.}~\bibnamefont {Rogal}},\ }\bibfield  {title} {\enquote {\bibinfo {title} {Efficient mapping of phase diagrams with conditional {Boltzmann} {Generators}},}\ }\href {\doibase 10.1088/2632-2153/ad849d} {\bibfield  {journal} {\bibinfo  {journal} {Machine Learning: Science and Technology}\ }\textbf {\bibinfo {volume} {5}},\ \bibinfo {pages} {045045} (\bibinfo {year} {2024})}\BibitemShut {NoStop}%
\bibitem [{\citenamefont {Wirnsberger}\ \emph {et~al.}(2022)\citenamefont {Wirnsberger}, \citenamefont {Papamakarios}, \citenamefont {Ibarz}, \citenamefont {Racanière}, \citenamefont {Ballard}, \citenamefont {Pritzel},\ and\ \citenamefont {Blundell}}]{wirnsberger_normalizing_2021}%
  \BibitemOpen
  \bibfield  {author} {\bibinfo {author} {\bibfnamefont {P.}~\bibnamefont {Wirnsberger}}, \bibinfo {author} {\bibfnamefont {G.}~\bibnamefont {Papamakarios}}, \bibinfo {author} {\bibfnamefont {B.}~\bibnamefont {Ibarz}}, \bibinfo {author} {\bibfnamefont {S.}~\bibnamefont {Racanière}}, \bibinfo {author} {\bibfnamefont {A.~J.}\ \bibnamefont {Ballard}}, \bibinfo {author} {\bibfnamefont {A.}~\bibnamefont {Pritzel}}, \ and\ \bibinfo {author} {\bibfnamefont {C.}~\bibnamefont {Blundell}},\ }\bibfield  {title} {\enquote {\bibinfo {title} {Normalizing flows for atomic solids},}\ }\href {\doibase 10.1088/2632-2153/ac6b16} {\bibfield  {journal} {\bibinfo  {journal} {Machine Learning: Science and Technology}\ }\textbf {\bibinfo {volume} {3}},\ \bibinfo {pages} {025009} (\bibinfo {year} {2022})}\BibitemShut {NoStop}%
\bibitem [{\citenamefont {Jung}, \citenamefont {Biroli},\ and\ \citenamefont {Berthier}(2024)}]{jung_normalizing_2024}%
  \BibitemOpen
  \bibfield  {author} {\bibinfo {author} {\bibfnamefont {G.}~\bibnamefont {Jung}}, \bibinfo {author} {\bibfnamefont {G.}~\bibnamefont {Biroli}}, \ and\ \bibinfo {author} {\bibfnamefont {L.}~\bibnamefont {Berthier}},\ }\bibfield  {title} {\enquote {\bibinfo {title} {Normalizing flows as an enhanced sampling method for atomistic supercooled liquids},}\ }\href {\doibase 10.1088/2632-2153/ad6ca0} {\bibfield  {journal} {\bibinfo  {journal} {Machine Learning: Science and Technology}\ }\textbf {\bibinfo {volume} {5}},\ \bibinfo {pages} {035053} (\bibinfo {year} {2024})}\BibitemShut {NoStop}%
\bibitem [{\citenamefont {Coretti}\ \emph {et~al.}(2025)\citenamefont {Coretti}, \citenamefont {Falkner}, \citenamefont {Geissler},\ and\ \citenamefont {Dellago}}]{coretti_learning_2025}%
  \BibitemOpen
  \bibfield  {author} {\bibinfo {author} {\bibfnamefont {A.}~\bibnamefont {Coretti}}, \bibinfo {author} {\bibfnamefont {S.}~\bibnamefont {Falkner}}, \bibinfo {author} {\bibfnamefont {P.~L.}\ \bibnamefont {Geissler}}, \ and\ \bibinfo {author} {\bibfnamefont {C.}~\bibnamefont {Dellago}},\ }\bibfield  {title} {{\selectlanguage {en}\enquote {\bibinfo {title} {Learning mappings between equilibrium states of liquid systems using normalizing flows},}\ }}\href {\doibase 10.1063/5.0253034} {\bibfield  {journal} {\bibinfo  {journal} {The Journal of Chemical Physics}\ }\textbf {\bibinfo {volume} {162}},\ \bibinfo {pages} {184102} (\bibinfo {year} {2025})}\BibitemShut {NoStop}%
\bibitem [{\citenamefont {Nicoli}\ \emph {et~al.}(2023)\citenamefont {Nicoli}, \citenamefont {Anders}, \citenamefont {Hartung}, \citenamefont {Jansen}, \citenamefont {Kessel},\ and\ \citenamefont {Nakajima}}]{nicoli_detecting_2023}%
  \BibitemOpen
  \bibfield  {author} {\bibinfo {author} {\bibfnamefont {K.~A.}\ \bibnamefont {Nicoli}}, \bibinfo {author} {\bibfnamefont {C.~J.}\ \bibnamefont {Anders}}, \bibinfo {author} {\bibfnamefont {T.}~\bibnamefont {Hartung}}, \bibinfo {author} {\bibfnamefont {K.}~\bibnamefont {Jansen}}, \bibinfo {author} {\bibfnamefont {P.}~\bibnamefont {Kessel}}, \ and\ \bibinfo {author} {\bibfnamefont {S.}~\bibnamefont {Nakajima}},\ }\bibfield  {title} {{\selectlanguage {en}\enquote {\bibinfo {title} {Detecting and mitigating mode-collapse for flow-based sampling of lattice field theories},}\ }}\href {\doibase 10.1103/PhysRevD.108.114501} {\bibfield  {journal} {\bibinfo  {journal} {Physical Review D}\ }\textbf {\bibinfo {volume} {108}},\ \bibinfo {pages} {114501} (\bibinfo {year} {2023})}\BibitemShut {NoStop}%
\bibitem [{\citenamefont {Kish}(1965)}]{kish_survey_1965}%
  \BibitemOpen
  \bibfield  {author} {\bibinfo {author} {\bibfnamefont {L.}~\bibnamefont {Kish}},\ }\href@noop {} {{\selectlanguage {English}\emph {\bibinfo {title} {Survey {Sampling}}}}}\ (\bibinfo  {publisher} {John Wiley \& Sons},\ \bibinfo {address} {New York},\ \bibinfo {year} {1965})\BibitemShut {NoStop}%
\bibitem [{\citenamefont {Shumailov}\ \emph {et~al.}(2024)\citenamefont {Shumailov}, \citenamefont {Shumaylov}, \citenamefont {Zhao}, \citenamefont {Papernot}, \citenamefont {Anderson},\ and\ \citenamefont {Gal}}]{shumailov_ai_2024}%
  \BibitemOpen
  \bibfield  {author} {\bibinfo {author} {\bibfnamefont {I.}~\bibnamefont {Shumailov}}, \bibinfo {author} {\bibfnamefont {Z.}~\bibnamefont {Shumaylov}}, \bibinfo {author} {\bibfnamefont {Y.}~\bibnamefont {Zhao}}, \bibinfo {author} {\bibfnamefont {N.}~\bibnamefont {Papernot}}, \bibinfo {author} {\bibfnamefont {R.}~\bibnamefont {Anderson}}, \ and\ \bibinfo {author} {\bibfnamefont {Y.}~\bibnamefont {Gal}},\ }\bibfield  {title} {{\selectlanguage {en}\enquote {\bibinfo {title} {{AI} models collapse when trained on recursively generated data},}\ }}\href {\doibase 10.1038/s41586-024-07566-y} {\bibfield  {journal} {\bibinfo  {journal} {Nature}\ }\textbf {\bibinfo {volume} {631}},\ \bibinfo {pages} {755--759} (\bibinfo {year} {2024})}\BibitemShut {NoStop}%
\bibitem [{\citenamefont {Vehtari}\ \emph {et~al.}(2024)\citenamefont {Vehtari}, \citenamefont {Simpson}, \citenamefont {Gelman}, \citenamefont {Yao},\ and\ \citenamefont {Gabry}}]{vehtari_pareto_2024}%
  \BibitemOpen
  \bibfield  {author} {\bibinfo {author} {\bibfnamefont {A.}~\bibnamefont {Vehtari}}, \bibinfo {author} {\bibfnamefont {D.}~\bibnamefont {Simpson}}, \bibinfo {author} {\bibfnamefont {A.}~\bibnamefont {Gelman}}, \bibinfo {author} {\bibfnamefont {Y.}~\bibnamefont {Yao}}, \ and\ \bibinfo {author} {\bibfnamefont {J.}~\bibnamefont {Gabry}},\ }\href {\doibase 10.48550/arXiv.1507.02646} {\enquote {\bibinfo {title} {Pareto {Smoothed} {Importance} {Sampling}},}\ } (\bibinfo {year} {2024}),\ \bibinfo {note} {arXiv:1507.02646 [stat.CO]}\BibitemShut {NoStop}%
\bibitem [{\citenamefont {Wolfe}\ \emph {et~al.}(1975)\citenamefont {Wolfe}, \citenamefont {Schlegel}, \citenamefont {Csizmadia},\ and\ \citenamefont {Bernardi}}]{wolfe_chemical_1975}%
  \BibitemOpen
  \bibfield  {author} {\bibinfo {author} {\bibfnamefont {S.}~\bibnamefont {Wolfe}}, \bibinfo {author} {\bibfnamefont {H.~B.}\ \bibnamefont {Schlegel}}, \bibinfo {author} {\bibfnamefont {I.~G.}\ \bibnamefont {Csizmadia}}, \ and\ \bibinfo {author} {\bibfnamefont {F.}~\bibnamefont {Bernardi}},\ }\bibfield  {title} {{\selectlanguage {en}\enquote {\bibinfo {title} {Chemical dynamics of symmetric and asymmetric reaction coordinates},}\ }}\href {\doibase 10.1021/ja00841a005} {\bibfield  {journal} {\bibinfo  {journal} {Journal of the American Chemical Society}\ }\textbf {\bibinfo {volume} {97}},\ \bibinfo {pages} {2020--2024} (\bibinfo {year} {1975})}\BibitemShut {NoStop}%
\bibitem [{\citenamefont {Quapp}(2005)}]{quapp_growing_2005}%
  \BibitemOpen
  \bibfield  {author} {\bibinfo {author} {\bibfnamefont {W.}~\bibnamefont {Quapp}},\ }\bibfield  {title} {{\selectlanguage {en}\enquote {\bibinfo {title} {A growing string method for the reaction pathway defined by a {Newton} trajectory},}\ }}\href {\doibase 10.1063/1.1885467} {\bibfield  {journal} {\bibinfo  {journal} {The Journal of Chemical Physics}\ }\textbf {\bibinfo {volume} {122}},\ \bibinfo {pages} {174106} (\bibinfo {year} {2005})}\BibitemShut {NoStop}%
\bibitem [{\citenamefont {Midgley}\ \emph {et~al.}(2023)\citenamefont {Midgley}, \citenamefont {Stimper}, \citenamefont {Simm}, \citenamefont {Schölkopf},\ and\ \citenamefont {Hernández-Lobato}}]{midgley_flow_2023}%
  \BibitemOpen
  \bibfield  {author} {\bibinfo {author} {\bibfnamefont {L.~I.}\ \bibnamefont {Midgley}}, \bibinfo {author} {\bibfnamefont {V.}~\bibnamefont {Stimper}}, \bibinfo {author} {\bibfnamefont {G.~N.~C.}\ \bibnamefont {Simm}}, \bibinfo {author} {\bibfnamefont {B.}~\bibnamefont {Schölkopf}}, \ and\ \bibinfo {author} {\bibfnamefont {J.~M.}\ \bibnamefont {Hernández-Lobato}},\ }\bibfield  {title} {\enquote {\bibinfo {title} {Flow {Annealed} {Importance} {Sampling} {Bootstrap}},}\ }in\ \href {https://openreview.net/forum?id=XCTVFJwS9LJ} {\emph {\bibinfo {booktitle} {The {Eleventh} {International} {Conference} on {Learning} {Representations}}}}\ (\bibinfo {year} {2023})\BibitemShut {NoStop}%
\bibitem [{\citenamefont {Schebek}, \citenamefont {Noé},\ and\ \citenamefont {Rogal}(2026)}]{schebek_scalable_2026}%
  \BibitemOpen
  \bibfield  {author} {\bibinfo {author} {\bibfnamefont {M.}~\bibnamefont {Schebek}}, \bibinfo {author} {\bibfnamefont {F.}~\bibnamefont {Noé}}, \ and\ \bibinfo {author} {\bibfnamefont {J.}~\bibnamefont {Rogal}},\ }\bibfield  {title} {{\selectlanguage {en}\enquote {\bibinfo {title} {Scalable {Boltzmann} generators for equilibrium sampling of large-scale materials},}\ }}\href {\doibase 10.1038/s41467-026-73900-9} {\bibfield  {journal} {\bibinfo  {journal} {Nature Communications}\ }\textbf {\bibinfo {volume} {17}},\ \bibinfo {pages} {5010} (\bibinfo {year} {2026})}\BibitemShut {NoStop}%
\bibitem [{\citenamefont {Durkan}\ \emph {et~al.}(2019)\citenamefont {Durkan}, \citenamefont {Bekasov}, \citenamefont {Murray},\ and\ \citenamefont {Papamakarios}}]{durkan_neural_2019}%
  \BibitemOpen
  \bibfield  {author} {\bibinfo {author} {\bibfnamefont {C.}~\bibnamefont {Durkan}}, \bibinfo {author} {\bibfnamefont {A.}~\bibnamefont {Bekasov}}, \bibinfo {author} {\bibfnamefont {I.}~\bibnamefont {Murray}}, \ and\ \bibinfo {author} {\bibfnamefont {G.}~\bibnamefont {Papamakarios}},\ }\href {\doibase 10.48550/ARXIV.1906.04032} {\enquote {\bibinfo {title} {Neural {Spline} {Flows}},}\ } (\bibinfo {year} {2019}),\ \bibinfo {note} {arXiv:1906.04032 [stat.ML]}\BibitemShut {NoStop}%
\bibitem [{\citenamefont {Rezende}\ \emph {et~al.}(2020)\citenamefont {Rezende}, \citenamefont {Papamakarios}, \citenamefont {Racaniere}, \citenamefont {Albergo}, \citenamefont {Kanwar}, \citenamefont {Shanahan},\ and\ \citenamefont {Cranmer}}]{rezende_normalizing_2020}%
  \BibitemOpen
  \bibfield  {author} {\bibinfo {author} {\bibfnamefont {D.~J.}\ \bibnamefont {Rezende}}, \bibinfo {author} {\bibfnamefont {G.}~\bibnamefont {Papamakarios}}, \bibinfo {author} {\bibfnamefont {S.}~\bibnamefont {Racaniere}}, \bibinfo {author} {\bibfnamefont {M.}~\bibnamefont {Albergo}}, \bibinfo {author} {\bibfnamefont {G.}~\bibnamefont {Kanwar}}, \bibinfo {author} {\bibfnamefont {P.}~\bibnamefont {Shanahan}}, \ and\ \bibinfo {author} {\bibfnamefont {K.}~\bibnamefont {Cranmer}},\ }\bibfield  {title} {\enquote {\bibinfo {title} {Normalizing {Flows} on {Tori} and {Spheres}},}\ }in\ \href {https://proceedings.mlr.press/v119/rezende20a.html} {\emph {\bibinfo {booktitle} {Proceedings of the 37th {International} {Conference} on {Machine} {Learning}}}},\ \bibinfo {series} {Proceedings of {Machine} {Learning} {Research}}, Vol.\ \bibinfo {volume} {119},\ \bibinfo {editor} {edited by\ \bibinfo {editor} {\bibfnamefont {H.~D.}\ \bibnamefont {III}}\ and\ \bibinfo {editor} {\bibfnamefont {A.}~\bibnamefont {Singh}}}\
  (\bibinfo  {publisher} {PMLR},\ \bibinfo {year} {2020})\ pp.\ \bibinfo {pages} {8083--8092}\BibitemShut {NoStop}%
\bibitem [{\citenamefont {Köhler}, \citenamefont {Krämer},\ and\ \citenamefont {Noe}(2021)}]{kohler_smooth_2021}%
  \BibitemOpen
  \bibfield  {author} {\bibinfo {author} {\bibfnamefont {J.}~\bibnamefont {Köhler}}, \bibinfo {author} {\bibfnamefont {A.}~\bibnamefont {Krämer}}, \ and\ \bibinfo {author} {\bibfnamefont {F.}~\bibnamefont {Noe}},\ }\bibfield  {title} {\enquote {\bibinfo {title} {Smooth {Normalizing} {Flows}},}\ }in\ \href {https://proceedings.neurips.cc/paper_files/paper/2021/file/167434fa6219316417cd4160c0c5e7d2-Paper.pdf} {\emph {\bibinfo {booktitle} {Advances in {Neural} {Information} {Processing} {Systems}}}},\ Vol.~\bibinfo {volume} {34},\ \bibinfo {editor} {edited by\ \bibinfo {editor} {\bibfnamefont {M.}~\bibnamefont {Ranzato}}, \bibinfo {editor} {\bibfnamefont {A.}~\bibnamefont {Beygelzimer}}, \bibinfo {editor} {\bibfnamefont {Y.}~\bibnamefont {Dauphin}}, \bibinfo {editor} {\bibfnamefont {P.~S.}\ \bibnamefont {Liang}}, \ and\ \bibinfo {editor} {\bibfnamefont {J.~W.}\ \bibnamefont {Vaughan}}}\ (\bibinfo  {publisher} {Curran Associates, Inc.},\ \bibinfo {year} {2021})\ pp.\ \bibinfo {pages} {2796--2809}\BibitemShut
  {NoStop}%
\bibitem [{\citenamefont {Tan}\ \emph {et~al.}(2025)\citenamefont {Tan}, \citenamefont {Bose}, \citenamefont {Lin}, \citenamefont {Klein}, \citenamefont {Bronstein},\ and\ \citenamefont {Tong}}]{tan_scalable_nodate}%
  \BibitemOpen
  \bibfield  {author} {\bibinfo {author} {\bibfnamefont {C.~B.}\ \bibnamefont {Tan}}, \bibinfo {author} {\bibfnamefont {A.~J.}\ \bibnamefont {Bose}}, \bibinfo {author} {\bibfnamefont {C.}~\bibnamefont {Lin}}, \bibinfo {author} {\bibfnamefont {L.}~\bibnamefont {Klein}}, \bibinfo {author} {\bibfnamefont {M.~M.}\ \bibnamefont {Bronstein}}, \ and\ \bibinfo {author} {\bibfnamefont {A.}~\bibnamefont {Tong}},\ }\href {\doibase 10.48550/arXiv.2502.18462} {\enquote {\bibinfo {title} {Scalable equilibrium sampling with sequential boltzmann generators},}\ } (\bibinfo {year} {2025}),\ \bibinfo {note} {arXiv:2502.18462 [cs.LG]}\BibitemShut {NoStop}%
\end{thebibliography}

\begin{thebibliography}{1}

\bibitem{wolfe_chemical_1975_supp}
Saul Wolfe, H.~Bernhard Schlegel, Imre~G. Csizmadia, and Fernando Bernardi.
\newblock Chemical dynamics of symmetric and asymmetric reaction coordinates.
\newblock {\em Journal of the American Chemical Society}, 97(8):2020--2024, April 1975.

\bibitem{quapp_growing_2005_supp}
Wolfgang Quapp.
\newblock A growing string method for the reaction pathway defined by a {Newton} trajectory.
\newblock {\em The Journal of Chemical Physics}, 122(17):174106, May 2005.

\bibitem{falkner_conditioning_2023_supp}
Sebastian Falkner, Alessandro Coretti, Salvatore Romano, Phillip~L Geissler, and Christoph Dellago.
\newblock Conditioning {Boltzmann} generators for rare event sampling.
\newblock {\em Machine Learning: Science and Technology}, 4(3):035050, September 2023.

\bibitem{noe_boltzmann_2019_supp}
Frank Noé, Simon Olsson, Jonas Köhler, and Hao Wu.
\newblock Boltzmann generators: {Sampling} equilibrium states of many-body systems with deep learning.
\newblock {\em Science}, 365(6457):eaaw1147, September 2019.

\bibitem{hummer_transition_2004_supp}
Gerhard Hummer.
\newblock From transition paths to transition states and rate coefficients.
\newblock {\em The Journal of Chemical Physics}, 120(2):516--523, January 2004.

\bibitem{e_transition-path_2010_supp}
Weinan E and Eric Vanden-Eijnden.
\newblock Transition-{Path} {Theory} and {Path}-{Finding} {Algorithms} for the {Study} of {Rare} {Events}.
\newblock {\em Annual Review of Physical Chemistry}, 61(1):391--420, March 2010.

\end{thebibliography}
\end{document}